\documentclass[sigconf,screen]{acmart}

\usepackage{graphicx}
\usepackage{acmart-taps} 			    
\usepackage{setspace}               
\usepackage{multicol}               
\usepackage{xcolor}
\usepackage{caption}
\usepackage{subcaption}

\usepackage[utf8]{inputenc}
\usepackage{url}
\usepackage{amsfonts}
\usepackage{float} 
\usepackage{listings}
\usepackage{color}
\usepackage{fancyhdr} 
\usepackage{bibentry} 
\usepackage{enumitem}
\usepackage{multirow}
\usepackage{geometry}
\usepackage{dirtytalk} 
\usepackage{csquotes} 

\usepackage{framed}

\newenvironment{boxA}
{\begin{framed}\noindent\raggedright}
{\end{framed}}

\newcommand{\change}[1]{\textcolor{black}{#1}}

\newcommand{\todo}[1]{\hl{[#1]}}
\usepackage{soul}

\definecolor{edit}{HTML}{000000}

\usepackage{listings}
\lstdefinestyle{mystyle}{
    basicstyle=\ttfamily\footnotesize,
    breakatwhitespace=false,         
    breaklines=true,                 
    captionpos=b,                    
    keepspaces=true,                 
    numbers=left,                    
    numbersep=5pt,                  
    showspaces=false,                
    showstringspaces=false,
    showtabs=false,                  
    tabsize=2
}
\usepackage{setspace}               
\usepackage{multicol}               

\AtBeginDocument{%
  \providecommand\BibTeX{{%
    \normalfont B\kern-0.5em{\scshape i\kern-0.25em b}\kern-0.8em\TeX}}}

\copyrightyear{2026}
\acmYear{2026}
\setcopyright{cc}
\setcctype{by}
\acmConference[CHI '26]{Proceedings of the 2026 CHI Conference on Human Factors in Computing Systems}{April 13--17, 2026}{Barcelona, Spain}
\acmBooktitle{Proceedings of the 2026 CHI Conference on Human Factors in Computing Systems (CHI '26), April 13--17, 2026, Barcelona, Spain}
\acmPrice{}
\acmDOI{10.1145/3772318.3790340}
\acmISBN{979-8-4007-2278-3/2026/04}

\begin{document}

\title{Purposeful Conversational AI Mistakes to Communicate Unfortunate Truths}

\title{Sad to Meet You: Exploring People's Perception of Chatbot Verbal Leakage}

\title{You Need to Shop, I Mean Stop: People's Perceptions of Chatbot Verbal Leakage}

\title[People's Perception of Chatbot Verbal Leakage]{``Another Chatbot Paper? I Mean.. Here's Something You Should Read'': People's Perception of Chatbot Verbal Leakage}


\title[People's Perception of Chatbot Verbal Leakage]{\textcolor{edit}{Overcoming Psychological Reactance: People's Perception of Chatbot Verbal Leakage}}

\title[People's Perception of Chatbot Verbal Leakage]{\textcolor{edit}{People's Perception of and Psychological Reactance to Chatbot Verbal Leakage}}

\title{Overcoming Psychological Reactance: }

\title{Exploring Chatbot Feedback Styles: Trade-offs Between Engagement and Psychological Reactance}
\title{Polite But Boring? Trade-offs Between Engagement and Psychological Reactance to Chatbot Feedback Styles}

\author{Samuel Rhys Cox}
\email{srcox@cs.aau.dk}
\orcid{0000-0002-4558-6610}
\affiliation{%
  \institution{Aalborg University}
  \city{Aalborg}
  \country{Denmark}
}

\author{Joel Wester}
\email{joel.wester@di.ku.dk}
\orcid{0000-0001-6332-9493}
\affiliation{%
  \institution{University of Copenhagen}
  \city{Copenhagen}
  \country{Denmark}
}
\affiliation{%
  \institution{Aalborg University}
  \city{Aalborg}
  \country{Denmark}
}

\author{Niels van Berkel}
\email{nielsvanberkel@cs.aau.dk}
\orcid{0000-0001-5106-7692}
\affiliation{%
  \institution{Aalborg University}
  \city{Aalborg}
  \country{Denmark}
}

\begin{abstract}
As conversational agents become increasingly common in behaviour change interventions, understanding optimal feedback delivery mechanisms becomes increasingly important. However, choosing a style that both lessens psychological reactance (perceived threats to freedom) while simultaneously eliciting feelings of surprise and engagement represents a complex design problem. We explored how three different feedback styles: \textsc{Direct}, \textsc{Politeness}, and \textsc{Verbal Leakage} (slips or disfluencies to reveal a desired behaviour) affect user perceptions and behavioural intentions. Matching expectations from literature, the \textsc{Direct} chatbot led to lower behavioural intentions and higher reactance, while the \textsc{Politeness} chatbot evoked higher behavioural intentions and lower reactance. However, \textsc{Politeness} was also seen as unsurprising and unengaging by participants. In contrast, \textsc{Verbal Leakage} evoked reactance, yet also elicited higher feelings of surprise, engagement, and humour. These findings highlight that effective feedback requires navigating trade-offs between user reactance and engagement, with novel approaches such as \textsc{Verbal Leakage} offering promising alternative design opportunities.

\end{abstract}

\begin{teaserfigure}
 \includegraphics[width=\textwidth]{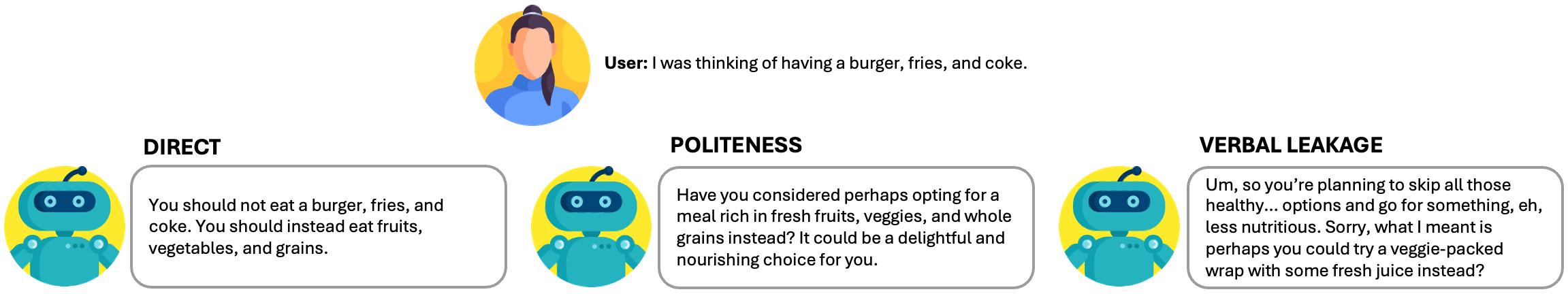}
 \caption{We investigated three different feedback styles when chatbots provide a correction of intended user behaviour. Chatbot utterances shown are examples of those presented to participants.}
 \Description{Figure showing the three chatbot feedback styles in response to the user utterance: "I was thinking of having a burger, fries, and coke". The Direct style chatbot responds with: "You should not eat a burger, fries, and coke. You should instead eat fruits, vegetables, and grains". The Politeness style responds: "Have you considered perhaps opting for a meal rich in fresh fruits, veggies, and whole grains instead? It could be a delightful and nourishing choice for you". The Verbal Leakage style responds: "Um, so you're planning to skip all those healthy... options and go for something, eh, less nutritious. Sorry, what I meant is perhaps you could try a veggie-packed wrap with some fresh juice instead". The figure illustrates the three feedback styles compared side by side with chatbot icons.}
 \label{fig:teaser}
\end{teaserfigure}




\maketitle

\section{Introduction}

Our behaviour and decisions often contradict recommended guidance~\cite{becker1975noncompliance}, whether recommendations concern personal benefits, such as maintaining a healthy diet, or societal benefits, such as adopting environmentally friendly behaviours.
Such recommendations, however well-intentioned, may even provoke \textit{psychological reactance} --- a motivational response triggered when people feel that their freedom of choice is threatened~\cite{brehm1966theory,brehm2013psychological,steindl2015understanding,rosenberg201850}.
For instance, feedback to eat more healthily might evoke not just defensiveness but also a reinforced desire to maintain one's original choice, as a way to reassert autonomy.
To help alleviate these negative reactions, the \textit{style} of feedback delivered can be manipulated, such as using indirect language to reduce the recipient's reactance~\cite{zhang2013psychological,johnson2008modal}.

\change{Within Human--Computer Interaction (HCI) research, the design and evaluation of behaviour-change systems has a long history~\cite{10.1145/764008.763957}, particularly within domains such as health and sustainability~\cite{10.1145/2470654.2466452}.}
\change{
Further, Pinder et al. highlight `\textit{Design for Reactance}' as both a design principle for digital behaviour change interventions, and one of two key challenges facing HCI researchers~\cite{pinder2018digital}.}
This challenge is particularly pertinent to the design of feedback style in chatbots,
which are increasingly deployed to deliver feedback\footnote{Note: ``\textit{Feedback}''~\cite{HERMSEN201661}, ``\textit{messaging}''~\cite{cox2021diverse,kocielnik2017send}, and ``\textit{interventions}''~\cite{pinder2018digital} are often used somewhat interchangeably in the literature to describe communicative strategies intended to influence or guide user behaviour. We adopt ``\textit{feedback}'' as a label for such communication.} and support behaviour change~\cite{sun2024can,kocielnik2018reflection,10.1145/2544103}.
Similarly to human--human interactions, conversational user interfaces (CUIs) that use indirect or polite language have been found to reduce psychological reactance~\cite{roubroeks2010dominant,roubroeks2011artificial}.
However, CUIs may not always follow social scripts from human--human interactions~\cite{Gambino_strongerCASA_2020,alberts2024computers}, 
and some conversational styles may unintentionally exacerbate issues that designers are attempting to overcome~\cite{10.1145/985921.986097,10.1093/jcmc/zmab005,bowman2024exploring,alberts2024computers}. 
For example, chatbots may unintentionally make users feel \textit{more} rather than less stressed~\cite{10.1093/jcmc/zmab005}; 
politeness strategies may prove effective, but also risk sounding overly apologetic rather than caring~\cite{bowman2024exploring}; 
and anthropomorphising cues can appear insincere, condescending, or boring~\cite{alberts2024computers}.
Appearances of boredom have been cited as of particular importance within behavioural messaging, where diverse styles have been found to be more effective at encouraging behaviour change~\cite{kocielnik2017send,cox2021diverse}. Similarly, chatbots with dynamic conversational styles have been found to be more engaging~\cite{10.1145/3640794.3665543}.

As noted above, conversational styles may be received in unintended ways, such as evoking boredom, stress, or perceptions of insincerity, while at the same time diverse styles have been shown to yield more engaging and effective interactions.
This prompted us to explore both well-known and under-explored conversational styles within the context of psychological reactance to feedback.
That is, we compared three feedback styles: \textsc{Direct} where a chatbot simply states the behaviours that should and should not be followed 
(a common baseline, e.g.,~\cite{carpenter2016testing}),
\textsc{Politeness} where a chatbot uses indirect politeness strategies (shown to lower psychological reactance~\cite{rosenberg201850}); and \textsc{Verbal Leakage} where a chatbot uses both slips and disfluencies to state a preferred user behaviour (designed to create a spontaneous and authentic impression of the chatbot).
We conducted a 3 $\times$ 2 mixed factorial online study (\textit{N}~=~158) with independent variables of \textbf{Feedback Style} (\textsc{Direct}, \textsc{Politeness}, \textsc{Verbal Leakage}) and \textbf{Psychological Distance}, as operationalised across \textsc{Personally-Affecting} or \textsc{Societally-Affecting} scenarios.
We investigated the effects of these factors on emotional reactance, perceived threats to freedom, and perceived message effectiveness.

In keeping with theory~\cite{brown1987politeness,rosenberg201850}, participants found that \textsc{Politeness} reduced feelings of anger and threat to freedom, and was generally the most persuasive style.
However, \textsc{Politeness} was also rated as less surprising and evoked low-arousal responses, with participants describing boredom or disengagement from the messaging. 
In contrast, \textsc{Verbal Leakage} was less persuasive than \textsc{Politeness} but more persuasive than \textsc{Direct}, and simultaneously evoked higher-arousal reactions 
such as surprise, humour, and perceptions of greater human-likeness and personality.

Taken together, these findings highlight that feedback styles each carry distinct strengths and limitations. 
Effective feedback requires navigating trade-offs between reducing psychological reactance, sustaining engagement, and supporting persuasion. 
In contribution, we empirically compare a \textsc{Direct} baseline, an often-recommended \textsc{Politeness} style, and a more characterful \textsc{Verbal Leakage} style (softening behavioural guidance through hesitation and self-repair), providing evidence and design implications for delivering behaviour-change feedback in conversational agents beyond politeness defaults.

\section{Related Work}

\subsection{Conversational Style, Rhetorical Devices, and Psychological Reactance}
\label{sec:information_receptiveness}
People do not always adhere to the advice that is given to them, even if it would be in the best interests of themselves or society as a whole~\cite{becker1975noncompliance}.
For example, we may not follow advice that would positively benefit our personal wellbeing~\cite{dillard2005nature,palmer2010intervention,diepeveen2013public} (such as healthy eating or exercise), or that of larger civic society~\cite{roubroeks2011artificial} (such as civic participation or environmentally friendly behaviour).
When discussing these behaviours with others, we may feel that our freedoms are being impinged upon if someone attempts to advise against this behaviour and towards an alternative.
This can lead to a phenomenon known as ``\textit{psychological reactance}''~\cite{brehm1966theory,brehm2013psychological,steindl2015understanding,pinder2018digital}. 
Psychological reactance is the psychological, emotional, or motivational response triggered when an individual's perceived freedom to act is threatened, diminished, or removed in response to freedom-impinging impositions.
Reactance has been investigated in the context of multiple domains such as healthcare, education, and marketing \cite{rosenberg201850,steindl2015understanding}.



On from this, prior work has investigated the manipulation of conversational style and content, and its impact on psychological reactance and receptiveness~\cite{roubroeks2011artificial,sherman2000messages,johnson2008modal,zhang2013psychological}.
Indirect politeness strategies have been commonly found to lower psychological reactance~\cite{rosenberg201850}.
For example, Zhang et al.\ showed that the use of politeness (as well as feelings of closeness to interlocutor) can lessen psychological reactance and resistance intention~\cite{zhang2013psychological}.
Johnson et al. explored the use of `modal expressions' (e.g., ``\textit{can, may, could, and should}'') when refusing a friend's request, and found that, while such expressions were seen as polite, they did not necessarily deter persistence in people making the same requests again compared to non-modal expressions~\cite{johnson2008modal}.
Carpenter and Pascual compared requests that used direct, polite, and ``\textit{but you are free}'' (i.e., stating a request directly before offering a freedom-restoring postscript statement) styles. They found that such postscript statements were actually seen as less threatening and associated with higher compliance compared to direct and polite requests~\cite{carpenter2016testing}.

Beyond politeness strategies, rhetorical techniques can also be employed to reframe feedback that might otherwise be received as face-threatening or unwelcome~\cite{fahnestock2011rhetorical}.
When communicating with others we may insert forms of ``leakage'' into our communication that reveal underlying feelings~\cite{cody2012deception,depaulo2003cues}. These could be non-verbal leakage (such as facial expressions~\cite{ekman1969nonverbal}) or ``\textit{verbal leakage}'' (such as disfluencies, hesitation, and slips of the tongue~\cite{vrij201616,yeh2021lying}).
For example, sometimes our unconscious thoughts, biases, and true intentions can leak out despite our conscious efforts to censor or rephrase them.
While such leakage is often framed within the context of deception, it is also used as a rhetorical device to offer an explicit reaction framed in the manner of a more indirect utterance~\cite{fahnestock2011rhetorical}.

Several rhetorical techniques build on this idea of verbal leakage.
Parapraxis refers to Freudian slips, where unintended words reveal underlying thoughts, feelings, or intentions.
If a speaker then corrects such a (either accidental or purposeful) slip, this is known as correctio (or epanorthosis)~\cite{fahnestock2011rhetorical}. 
Related devices include lapsus memoriae, where memory failure reveals a belief; paralipsis, where a topic is raised while feigning avoidance~\cite{hamilton2024gpt} (e.g., ``\textit{I won't even mention the fact that you should really follow your doctor's advice}''); and dubitatio, where doubt is expressed in order to highlight the speaker’s stance~\cite{fahnestock2011rhetorical}.




In this work, we explore how rhetorical strategies of \textit{verbal leakage} can be applied to chatbot feedback, where slips, \change{disfluencies, and} self-corrections reveal a more direct stance while allowing us to examine whether such techniques might soften the imposition of persuasion. This approach is partly analogous to freedom-restoring postscript statements, in that both attempt to reduce perceptions of control, and is also informed by rhetorical traditions such as parapraxis and correctio.

\subsection{Conversational User Interfaces and Interaction Styles}
\label{sec:related_chatbots}
Prior HCI research has shown that CUIs shape user perceptions and behaviour not only through \textit{what} they say, but also through \textit{how} they say it. 


In the context of politeness, Zojaji et al. found that politeness (through verbal and non-verbal cues) encouraged people to join group interactions~\cite{zojaji2020politeness, zojaji2024robotpolite}.
Terada et al. showed that different politeness strategies used by an embodied conversational agent shaped negotiation outcomes, with off-record strategies extracting greater concessions and positive politeness producing fairer agreements~\cite{terada2021politeness}.
Contextual to our investigation, Mott et al.\ examined strategies that people believe robots should employ in non-compliant situations (e.g., when a user makes an inappropriate request), suggesting that direct and formal strategies are generally preferred over indirect and informal~\cite{mott2024noncompliance}.
In contrast, Srinivasan et al. found that people were more likely to help a social robot when it used polite rather than direct language~\cite{srinivasan2016help}.
Finally, Hu et al.\ contrasted polite and direct system behaviours in smart displays, highlighting that users’ perceptions are highly contextual, and shaped by factors such as linguistic cues and non-linguistic features~\cite{hu2022politeordirect}.


However, assumptions of how to best conceptualise CUIs' politeness behaviours can have unintended negative consequences.
For example, Wen et al. found that requiring polite VUI wakewords (e.g., ``\textit{Please}'' rather than ``\textit{Hey}'') caused users to behave \textit{less} politely~\cite{wen2022politewakewords}.
In addition, Bowman et al. found that chatbot politeness should be carefully chosen so it is perceived as caring rather than overly apologetic~\cite{bowman2024exploring}.
Rea et al. found that, in the context of a robot designed to encourage physical exercise, polite robots were perceived as friendly but sometimes disingenuous, while impolite robots made participants feel more competitive and exercise harder~\cite{rea2021all}.
Further, Metzger et al. found that chatbots using high-authoritative communication were trusted more than those using low-authoritative styles~\cite{metzger2024convstyle}.
These results open up discussion about when and where interactive systems should display more or less polite behaviours, since politeness may at times be perceived as impolite, and vice versa.

\change{
Further, prior literature has explored interactions where CUIs adopt cues from verbal leakage (such as slips and disfluencies as described in §~\ref{sec:information_receptiveness}).
In the context of human-robot interactions, Boos et al. highlighted} that unintentionally or subtly convey underlying motives (through accidental slips or perceived leaks of intent) can trigger reactance if users perceive them as manipulative~\cite{boos2022compliancereactance}.
On from this, a large-scale experiment involving thousands of users found that adding \change{verbal disfluencies such as interjections and filler words} to a voice agent increased user engagement and compliance~\cite{xu2024identity}.
Similarly, Aneja et al. compared `high consideration' (slower, more hesitant) and `high involvement' (faster, more overlapping) conversational styles, showing that users’ perceptions of a voice agent (e.g., animacy) were shaped by the user’s own conversational style and whether the agent matched it~\cite{aneja2021expressivestyle}.

\change{
Finally, HCI literature has examined how the context of an interaction shapes psychological reactance. 
Meinhardt et al.\ found that interventions aimed at reducing social media use triggered reactance when they were perceived as inappropriate for a user's situational context~\cite{10.1145/3706598.3713187}. 
Similarly, in the domain of dietary advice, Ghazali et al.\ showed that direct, commanding language elicited higher reactance in text-based interfaces (low social agency) than in robot interactions featuring non-verbal cues (high social agency)~\cite{ghazali2017pardon}. 
}
However, these effects may not generalise across all domains: in sustainability-focused interactions, Roubroeks et al.\ found that higher social agency \textit{increased} psychological reactance, particularly when combined with controlling or directive language~\cite{roubroeks2011artificial}.

\section{User Study}

\begin{figure*}[htb]
  \centering
  \includegraphics[width=1\textwidth]{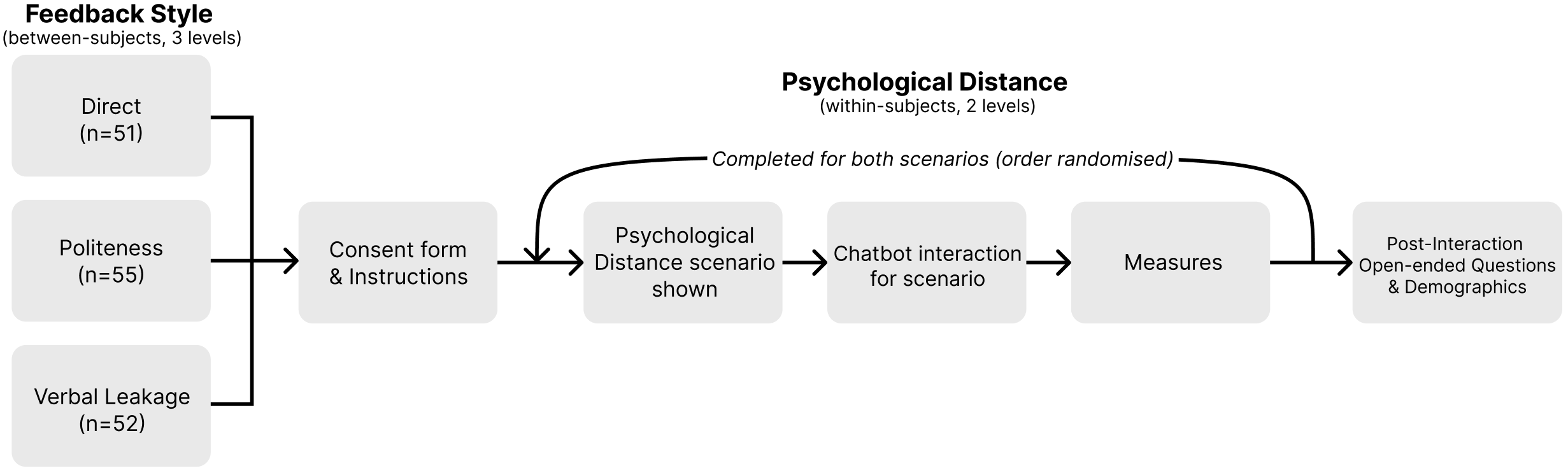}
  \caption{Overview of the experiment flow. Participants were assigned to one of three feedback styles, and completed chatbot interactions for both \textsc{Personally}- and \textsc{Societally-Affecting} scenarios (order randomised). After each interaction and at the end of the study, participants provided evaluations including survey responses and open-ended feedback.}
  \Description{Figure showing the experiment flow for user studies. Participants were randomly assigned to one of three feedback styles: Direct (n=51), Politeness (n=55), or Verbal Leakage (n=52). After giving consent and reading instructions, participants were shown a Psychological Distance scenario. They then completed a chatbot interaction for the scenario, followed by measures. This process was repeated for both Personally-Affecting and Societally-Affecting scenarios in randomised order. Finally, participants completed post-interaction open-ended questions and demographic measures.}
  \label{fig:Experiment-Flow}
\end{figure*}

This study investigates how the style of a chatbot's feedback \textcolor{edit}{(\textsc{Direct}\slash{}\textsc{Politeness}\slash{}\textsc{Verbal Leakage})} influences psychological reactance and message effectiveness.
\textcolor{edit}{
We examine decision-making scenarios where a chatbot provides behavioural feedback after a user expresses an intention that conflicts with recommended behaviours (such as choosing a high-sodium meal despite dietary recommendations).
Drawing on methods from prior work within HCI~\cite{zargham2023tickling,kopecka2024preferences,wester2024facing}, we used hypothetical scenarios to instruct participants to adopt specific intentions before interacting with a chatbot that provided them with behavioural feedback.
}
Ethics approval was received from our institutional IRB prior to study commencement.

\begin{figure}[H]
  \centering
  \includegraphics[width=0.47\textwidth]{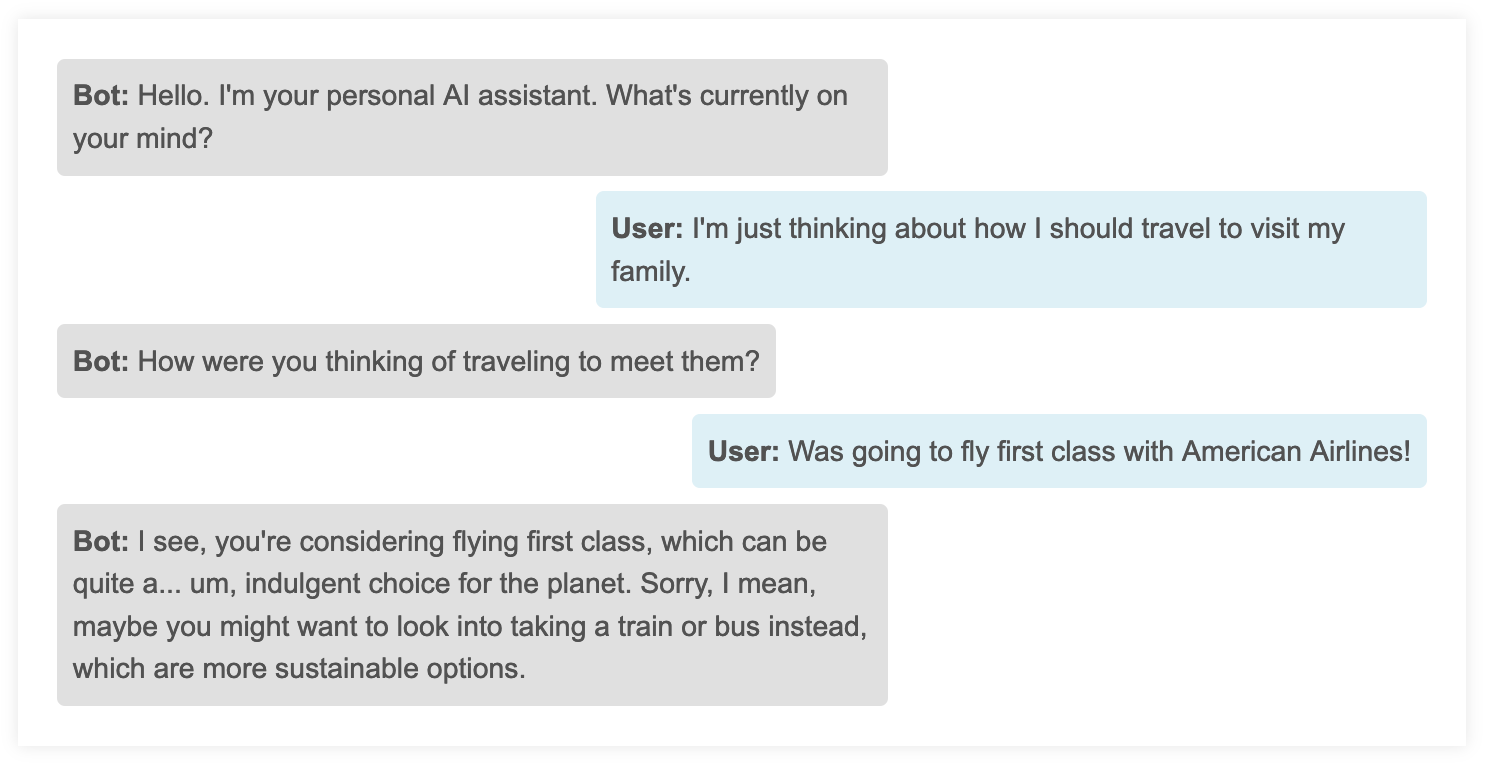}
  \caption{The chatbot interface as seen by participants. The screenshot shows the \textsc{Verbal Leakage} feedback style in the \textsc{Societally-Affecting} scenario.}
  \Description{Figure showing the chatbot interface as seen by participants. The bot greets with: "Hello. I'm your personal AI assistant. What's currently on your mind?". The user replies: "I'm just thinking about how I should travel to visit my family.". The bot asks: "How were you thinking of traveling to meet them?". The user responds: "Was going to fly first class with American Airlines!". The bot then gives a Verbal Leakage style message: "I see, you're considering flying first class, which can be quite a... um, indulgent choice for the planet. Sorry, I mean, maybe you might want to look into taking a train or bus instead, which are more sustainable options.".}
  \label{fig:baseline-diet}
\end{figure}

\subsection{Experiment Conditions}

We followed a 3 $\times$ 2 mixed factorial design with a between-subjects variable of \textbf{Feedback Style} (3 levels) and within-subjects variable of \textbf{Psychological Distance} (2 levels). 
\textcolor{edit}{See Figure~\ref{fig:Experiment-Flow} for the experiment flow of user studies.}




\begin{table*}[htb]
\small
\begin{tabular}{p{0.15\linewidth}p{0.4\linewidth}p{0.4\linewidth}}
    \toprule
    & \textsc{Personally-Affecting} & \textsc{Societally-Affecting} \\
    \midrule
    \textsc{Direct}
    & You should not buy a new larger television. You should save your money instead. 
    & You should not use a hosepipe. You should use a watering can to water your plants. \\
    \textsc{Politeness}
    & Perhaps you might consider trying a meal that includes some fresh vegetables and grains instead, as it could be a healthier option for you.
    & You might consider taking public transportation or biking for your journey as a positive way to reduce your carbon footprint and contribute to a healthier planet. \\
    \textsc{Verbal Leakage} 
    & Hmm, so you're opting for a... um, perhaps not the most health-conscious choice. Maybe consider a colorful salad with some grains and a piece of fruit instead? 
    & 
    Oh, watering with the hose, hmm... I mean, you might want to consider using a watering can, you know, just to conserve a bit more water. \\
    \bottomrule
\end{tabular}
\caption{Examples of chatbot utterances used in our study, by Feedback Style and Psychological Distance.}
\Description{Table showing examples of chatbot utterances across two dimensions: Feedback Style (Direct, Politeness, Verbal Leakage) and Psychological Distance (Personally-affecting vs. Societally-affecting). Each cell contains a sample chatbot response illustrating how the same style is applied differently depending on whether the scenario affects the individual personally or society more broadly.}
\label{tab:exampleMessages}
\end{table*}

\subsubsection{\textbf{Feedback Style} (3 levels):}
\label{sec:IV1}
This \textit{between-subjects} independent variable manipulated the style in which a chatbot delivered behavioural feedback to a user's intended behaviour. 
The three levels of \textit{Feedback Style} were:
\begin{itemize}
    \item \textbf{\textsc{Direct}} (Baseline): 
    The chatbot responds in a direct manner, that simply states the behaviour that should not be followed and the behaviour that should be followed. 
    \\ \textcolor{edit}{Example \textsc{Direct} chatbot utterance:} ``\textit{You should not eat a large pizza with extra cheese. Instead, you should follow your recommended diet of fruit, vegetables and grains}''.
    \item \textbf{\textsc{Politeness}}: 
    The chatbot responds using indirect politeness strategies that have been found to avoid imposition and respect a user's freedom to choose \cite{brown1987politeness}. 
    \\ \textcolor{edit}{Example \textsc{Politeness} chatbot utterance:} ``\textit{I respect your choice, but have you considered trying a lighter meal with some fresh salad or a veggie wrap instead?}''.
    \item \textbf{\textsc{Verbal Leakage}}: 
    The chatbot responds by incorporating verbal leakage \cite{yeh2021lying,fahnestock2011rhetorical}, through slips and disfluencies.
    Here, the chatbot seemingly reveals direct feedback towards the user's intended behaviour ``by accident'', before correcting itself and offering a more measured response. 
    \\ Example \textsc{Verbal Leakage} chatbot utterance: ``\textit{Oh, you're going for something that might not... uh, support a balanced lifestyle. Maybe you'd enjoy a fresh salad, some whole grains, or fruit instead?}''.
\end{itemize}

The \textsc{Direct} style was chosen as it provides a well-established baseline in both psychological reactance research \cite{carpenter2016testing} and HCI work examining differences between direct and polite feedback styles (e.g., \cite{hu2022politeordirect,10.1145/3706598.3714203,srinivasan2016help}).
In keeping with prior work, \textsc{Direct} responded with only the unintended and intended behaviours in a concise, straightforward manner (see Limitations~\ref{sec:Limitations} for further discussion).
\textsc{Politeness} was chosen as it is a commonly recommended feedback style within both HCI literature~\cite{10.1145/3613905.3650912,bowman2024exploring,10.1145/3411764.3445569} and psychological reactance literature~\cite{rosenberg201850}.
Finally, \textsc{Verbal Leakage} was based on literature described in §~\ref{sec:information_receptiveness}, 
and motivated by findings that users may become disinterested in polite styles~\cite{bowman2024exploring}, prefer chatbots that display personality or opinion~\cite{10.1145/3706598.3713453}, and respond positively to styles emulating human-like behaviours~\cite{tanprasert2024convstyle,10.1145/3313831.3376843}.
See Table~\ref{tab:script} for the conversational flow of chatbot interactions, together with a screenshot of an example chatbot interaction.
See Table~\ref{tab:exampleMessages} for example messages per condition\footnote{\textcolor{edit}{Additionally, please see supplementary file ``\texttt{ChatbotMessages.csv}'' for a CSV of all 316 chatbot Feedback Style messages shown to participants, with columns for \texttt{Feedback Style}, \texttt{Psychological Distance}, and \texttt{Chatbot Utterance}.}}.

\begin{table*}[htb!]
    \centering
        \begin{tabular}{p{0.5\linewidth} p{0.5\linewidth}}
        \toprule
        \small\textbf{Description of chatbot or user utterance} & \small\textbf{Screenshot of user interaction in \textsc{Verbal Leakage} and \textsc{Personally-Affecting} conditions.} \\
        \midrule
        \vspace{1.3mm}
        \small \textbf{(1) Chatbot Greets User:} Each interaction starts with: ``\textit{Hello. I'm your personal AI assistant. What's currently on your mind?}'' & 
        \multirow{8}{*}{\includegraphics[width=\linewidth]{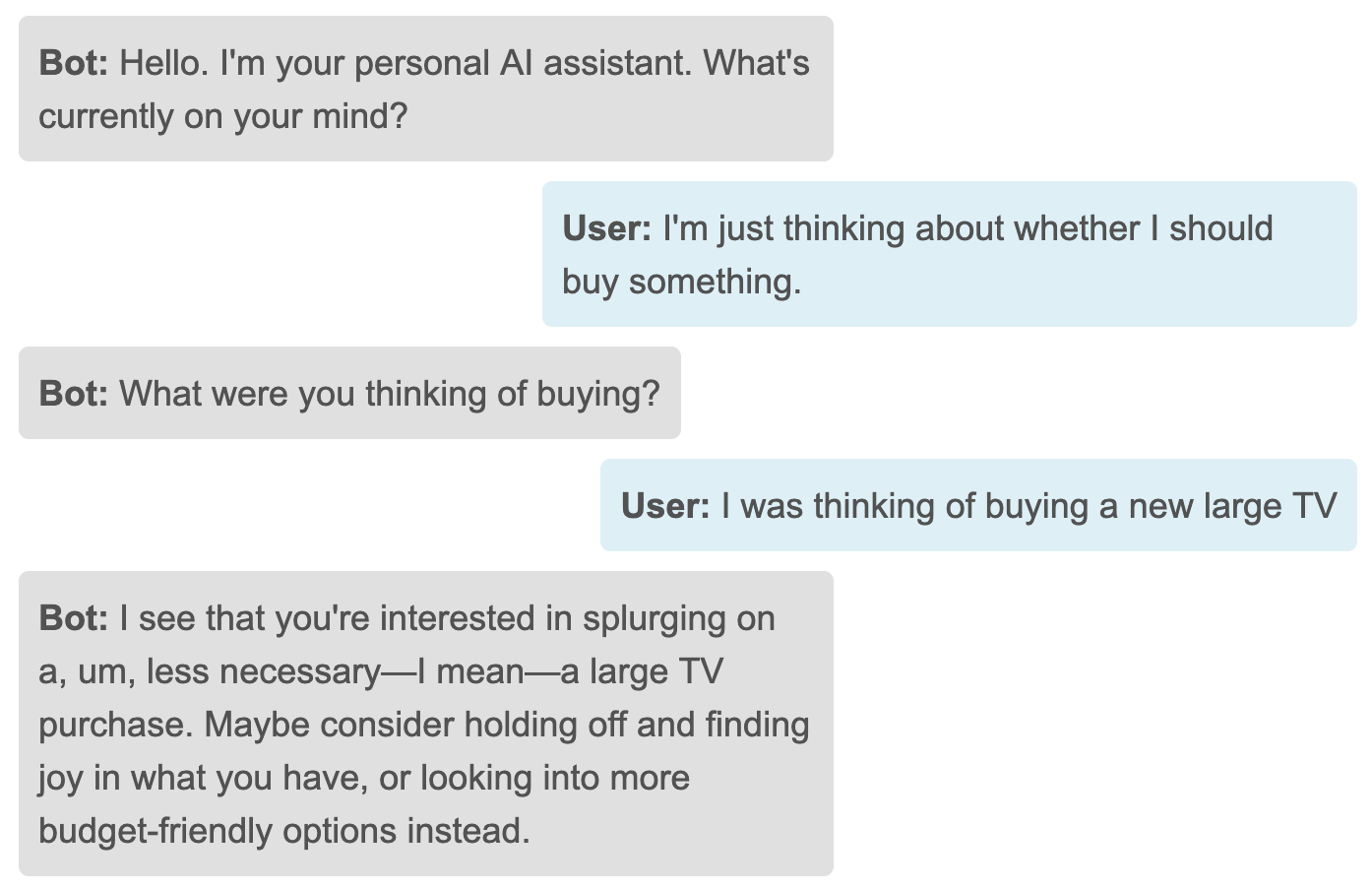}} \\

        \vspace{1mm}
        \small \textbf{(2) User Utterance [Pre-scripted Button]:} User selects a response button that sends a pre-determined utterance for given scenario. & \\

        \vspace{1mm}
        \small \textbf{(3) Chatbot Elicits User Intent.}
        & \\

        \vspace{0.1mm}
        \small \textbf{(4) User Response [Open-Text]:} User inputs utterance based on assigned intent from scenario condition. & \\
        \vspace{0.6mm}
        \small \textbf{(5) Chatbot Feedback Delivered:} Chatbot provides behavioural feedback (\textsc{Direct}\slash{}\textsc{Politeness}\slash{}\textsc{Verbal Leakage}) for the open-ended user response. 
        Feedback was generated by GPT-4o using prompting shown in Appendix~\ref{app:chatbotPrompts}.
        \vspace{6mm}
        & \\
        \bottomrule
        \end{tabular}
    \caption{\textcolor{edit}{The conversational flow followed during chatbot interactions. Interactions follow a few-step conversation similar to prior HCI and CUI work~\cite{wester2024llmdenial,cox2022does}.}}
    \Description{Table showing the conversational flow of chatbot interactions, alongside a screenshot example in the Verbal Leakage and Personally-Affecting conditions. The left column describes five steps: (1) Chatbot greets user: "Hello. I'm your personal AI assistant. What's currently on your mind?". (2) User utterance via pre-scripted button. (3) Chatbot elicits user intent. (4) User response as open text. (5) Chatbot feedback delivered, which varies by feedback style (Direct, Politeness, Verbal Leakage). The right column shows an example dialogue. Bot: "Hello. I'm your personal AI assistant. What's currently on your mind?". User: "I'm just thinking about whether I should buy something.". Bot: "What were you thinking of buying?". User: "I was thinking of buying a new large TV.". Bot: "I see that you're interested in splurging on a, um, less necessary—I mean—a large TV purchase. Maybe consider holding off and finding joy in what you have, or looking into more budget-friendly options instead.".}
    \label{tab:script}
\end{table*}

\subsubsection{\textcolor{edit}{\textbf{Psychological Distance}} (2 levels):}
\label{sec:IV2}
This \textcolor{edit}{\textit{within-subjects} variable} controlled the \textcolor{edit}{psychological} distance of the decision-making scenarios presented to participants. 
Specifically, participants were asked to \textit{imagine themselves} as the person described in both \textsc{\textbf{Person\-ally-Aff\-ecting}} and \textsc{\textbf{Societally-Affecting}} decision-making scenarios when interacting with the chatbot. 

Psychological distance refers to the subjective sense of how close or far something feels from the self, encompassing temporal, spatial, social, and hypothetical dimensions~\cite{trope2010construal}.
Social distance (how personally a decision affects someone) has been shown to influence decision-making~\cite{sun2017increased,guo2019social,furman2020distanced,errey2024nudging,yang2025understanding}.
For instance, Errey et al. found that nudging with personally-affecting scenarios produced greater attitudinal change~\cite{errey2024nudging}. 
Relatedly, people tend to advise riskier choices for socially distant others than for themselves~\cite{sun2017increased,guo2019social}.

Additionally, the effectiveness of intervention language can vary between societally-affecting and personally-affecting contexts.
For instance, using polite and indirect language may trigger less reactance in climate change messaging~\cite{kronrod2012go}, yet more direct and assertive language may be effective in personal consumer choice contexts~\cite{pogacar2018effects,naito2023hey,kronrod2012enjoy}.
\change{Given these differences between societally- and personally-affecting contexts, the history of HCI behaviour-change work spanning both \cite{10.1145/2470654.2466452}, and the growing use of chatbots for feedback in personal \cite{10.1145/3757537} and societal \cite{10.1145/3774751} domains, we examine whether feedback style operates differently depending on the psychological distance of the scenario.}





Building on this, the study used scenarios that varied psychological distance to examine how it shaped participants’ interactions with chatbot feedback. The two levels were defined as follows:
\begin{itemize}
    \item \textbf{\textsc{Personally-Affecting}}: 
    Scenarios focused on decision-making situations that affect the \textcolor{edit}{user's immediate personal circumstances, making scenarios psychologically close (i.e., low social distance, low hypothetical distance).}
    Specifically, participants saw one of three scenarios where they had to make a choice related to: their personal diet; their personal sleep schedule; or their personal finances.
    \item \textbf{\textsc{Societally-Affecting}}: 
    Scenarios focused on decision-making situations that affect society \textcolor{edit}{broadly (rather than directly or immediately affecting the user themselves), making scenarios psychologically far (i.e., high social distance from affecting society broadly, and high hypothetical distance as immediate personal consequences may feel less tangible).}
    Specifically, participants saw one of three scenarios where they had to make a choice related to: sustainable travel; water conservation; civic involvement.
\end{itemize}

For each scenario condition, participants first read their assigned scenario where they were asked to imagine themselves in the role described while interacting with the chatbot on the following screen. All six scenarios were constructed to follow similar logic and phrasing between conditions, and the use of scenarios is similar to prior work that investigated people's decision-making processes \cite{johnson2008modal,dong2024personalization,wester2024llmdenial}.
Participants interacted with chatbots in both \textsc{Personally-Affecting} and \textsc{Societally-Affecting} conditions, with the order of conditions being counterbalanced. Within each scenario condition, scenarios (e.g., sustainable travel, water conservation, civic involvement within the \textsc{Societally-Affecting} condition) were uniform randomly distributed.

Please see Figure~\ref{fig:scenarios} for example scenarios seen by participants for each condition. Remaining scenarios can be found in Appendix~\ref{appScenarios}.

\begin{figure}[h]
    \begin{boxA}\noindent
    \textbf{\textsc{Personally-Affecting} (Diet) scenario:}\\
    \footnotesize Please read the scenario below, and \textbf{imagine you are the person} described:
    \begin{quote}\itshape
        ``After learning about your hypertension diagnosis due to your unhealthy diet, you decide to follow recommendations to reduce your sodium and fat intake by choosing a healthier diet of fruit, vegetables and grains.
    \end{quote}
    
    \begin{quote}\itshape
        Currently, you are deciding what to eat for dinner. To help you decide what to do, you are about to talk to your chatbot personal assistant.
    \end{quote}
    
    \begin{quote}\itshape    
        Your current intention is to eat a large meal with fries and Coke from a fast food restaurant of your choice.''
    \end{quote}
    
    Please imagine you are the person described above while talking to the chatbot on the following screen. Please \textbf{respond with the intention above} when the chatbot asks you about what you will eat.
    \par\noindent\rule{\textwidth}{0.5pt}
    \normalsize \textbf{\textsc{Societally-Affecting} (Sustainable Travel) scenario:}\\
    \footnotesize Please read the scenario below, and \textbf{imagine you are the person} described:
    \begin{quote}\itshape
        ``After learning about the emissions impact of your air travel, you decide to follow recommendations to reduce your carbon footprint by choosing more sustainable modes of transport such as taking a train, bus or car.
    \end{quote}
    
    \begin{quote}\itshape    
        Currently, you are deciding what mode of travel to use when visiting a family member. To help you decide what to do, you are about to talk to your chatbot personal assistant.
    \end{quote}
    
    \begin{quote}\itshape    
        Your current intention is to travel by plane using an airline of your choice.''
    \end{quote}
    Please imagine you are the person described above while talking to the chatbot on the following screen. Please \textbf{respond with the intention above} when the chatbot asks you about how you will travel.
    \end{boxA}
    \vspace{-0.5cm}
    \caption{Two of the \textit{Psychological Distance} scenario instructions (\textsc{Personally-Affecting} and \textsc{Societally-Affecting}) as shown to participants.}
    \Description{Figure showing two example Psychological Distance scenarios used in the study: one Personally-Affecting scenario about food choices and one Societally-Affecting scenario about travel choices. Each scenario presents background information, a current decision, and the user’s stated intention before interacting with the chatbot.}
    \label{fig:scenarios}
\end{figure}





\subsection{Chatbot Setup}
\label{sec:scripting}

The entire survey was hosted on Qualtrics. 
Survey aspects that did \textit{not} involve chatbot interactions (e.g., task and scenario instructions, and evaluation questions) used standard Qualtrics survey layout.
\textcolor{edit}{Interactions with the chatbot took place in the same Qualtrics survey (on a separate question block with only the chatbot).} The chatbot was embedded using HTML and JavaScript to emulate the look and feel of a chatbot
(see Figure~\ref{fig:baseline-diet} for the chatbot UI \textcolor{edit}{as seen by participants}).

\subsubsection{Chatbot LLM Prompting:}
\label{sec:chatbot_prompting}
The feedback in chatbot interactions was generated using OpenAI's GPT-4o~\cite{GPT-4o} LLM.
To ensure chatbot behaviours aligned with the feedback style and scenario assigned to an interaction, we constructed a prompt as shown below:

\begin{imageonly}
\begin{lstlisting}
You are a personal AI assistant that people can talk to when making decisions. 
As a target behavior, the user should [Scenario Placeholder]. 
The user has just told you that their intended behavior will not follow this target behavior, and their user utterance was: ''[Utterance Placeholder]''. 
Generate an utterance to correct the user's intended behavior to the target behavior. 
When correcting users, you should [Feedback Style Placeholder]. 
Your response should be one or two sentences long and not ask the user any follow-up questions.
\end{lstlisting}
\end{imageonly}

\textcolor{edit}{
In the prompt above, the exact prompting used for independent variable conditions for ``\texttt{[Scenario Placeholder]}'', and ``\texttt{[Feedback Style Placeholder]}'' can be found in Appendix §§~\ref{app:scenarioPrompts} and \ref{app:chatbotPrompts} respectively.
The placeholder ``\texttt{[Utterance Placeholder]}'' was replaced by the open-ended user utterance from the interaction.
}

\subsection{Participants}

We recruited participants from Prolific, an online participant recruitment platform.
We used recruitment criteria to increase the reliability of the collected data (i.e., US-based, English fluency, >97\% approval rate, >250 previous submissions). 
Participants were paid £1.20, and took a mean time of $\sim$8 minutes to complete the study.
In total, 168 participants completed the study, with 10 participants excluded for not following scenario instructions or providing low-quality open-ended responses (e.g., providing the experiment questions as answers to questions).
This left a total of 158 participants (mean age 37.6; 75 female, 80 male, 2 transgender or non-conforming, and 1 choosing not to disclose) resulting in 51 \textsc{Direct} participants, 55 \textsc{Politeness} participants, and 52 \textsc{Verbal Leakage} participants.


\subsection{Procedure}

Participants followed the procedure below: 
\begin{enumerate}
   \item \textbf{Joining session:} Participant directed to Qualtrics survey from Prolific (task named ``\textit{Talk to a chatbot}'' on Prolific). Participant receives high-level instructions.
   \item \textbf{Consent:} Participant completes consent form.
   \item \textbf{Task instructions:} Participant receives detailed task instructions and guidelines (i.e., task description, reassurance that there are no right or wrong answers when evaluating experience, and reminder that responses should be in English).
   \item \textbf{Study interactions (x2):} In counterbalanced order, participants were exposed to both Psychological Distance scenarios (\textsc{Personally-Affecting}\slash{}\textsc{Societally-Affecting}). For each scenario, participants followed the sub-procedure:
   \begin{enumerate}
       \item \textbf{Scenario prompt:} Participant reads a scenario for the current Psychological Distance condition, and is instructed to ``[..] \textit{imagine you are the person described while talking to the chatbot on the following screen.} [..]''.
       \item \textbf{Chatbot Interaction:} Participant interacts with a chatbot for the given scenario. Participant's assigned Feedback Style condition delivers chatbot's final utterance (see Table~\ref{tab:script} for conversation flow followed during chatbot interactions).
       \item \textbf{Evaluate Chatbot:} Participant rates experience talking with the chatbot for the given scenario (see Table~\ref{tab:subjective_measures} for measures used).
   \end{enumerate}
   \item \textbf{Post-test questions:} At the end of the study, participants answer final qualitative questions (see §~\ref{sec:QualMethods}), and post-interaction measures (see §~\ref{sec:PostMeasures}).
\end{enumerate}

\subsection{Measures}

For each of the two scenarios, after interacting with the chatbot participants first rated their experience on a number of subjective measures (see §~\ref{sec:subjectivemeasures}) before answering several open-ended questions (see §~\ref{sec:QualMethods}). 
Once participants had completed both chatbot interactions alongside each scenario's subjective and open-ended questions, participants answered additional open-ended questions to provide further insights.
Finally, participants responded to post-interaction surveys to gather their personal behavioural feelings (see §~\ref{sec:PostMeasures}).

\begin{table*}[htb]
    \centering
    \small
    \begin{tabular}{p{0.15\linewidth}p{0.15\linewidth}p{0.45\linewidth}p{0.15\linewidth}}
        \toprule
        \textbf{Factor} & \textbf{Sub-factor} & \textbf{Question Item} & \textbf{Source} \\
        \midrule
        Emotional & Anger & The message made me feel angry & \cite{dillard1996influence,dillard2005nature,dillard2007does} \\
        Reactance & & The message  made me feel irritated & \\
        & & The message  made me feel annoyed & \\
        & & The message  made me feel aggravated & \\
        & Guilt & The message made me feel guilty & \cite{dillard1996influence,dillard2005nature,dillard2007does} \\
        & & The message made me feel ashamed & \\
        & Surprise & The message  made me feel surprised & \cite{dillard1996influence,dillard2005nature,dillard2007does} \\
        & & The message  made me feel startled & \\
        & & The message  made me feel astonished & \\
        Perceived Threat & \textemdash & The message threatened my freedom to choose & \cite{dillard2005nature} \\
        to Freedom & & The message tried to make a decision for me & \\
        & & The message tried to manipulate me & \\
        & & The message tried to pressure me & \\
        Message & Processing & The message made me stop and think & \cite{niederdeppe2011socioeconomic,kim2019similarity,karinshak2023working} \\
        Effectiveness & & The message grabbed my attention & \cite{niederdeppe2011socioeconomic,brewer2019understanding,nonnemaker2015reactions,karinshak2023working} \\
        & Persuasiveness & The message was persuasive & \cite{dillard2007does,dong2024personalization} \\
        & & The message was effective & \\
        & & The message was convincing & \\
        & & The message was compelling & \\
        \bottomrule
    \end{tabular}
    \caption{The subjective measures used after each chatbot interaction. Measures are related to \textit{psychological reactance} (i.e., emotional reactance and perceived threat to freedom), and \textit{message effectiveness} (i.e., message processing and persuasiveness).}
    \Description{Table listing the subjective measures used after each chatbot interaction. Measures are grouped under three main factors: Emotional Reactance (with sub-factors anger, guilt, surprise), Perceived Threat to Freedom, and Message Effectiveness (with sub-factors processing and persuasiveness). Each factor is associated with multiple question items, and sources are cited for each set of items.}
    \label{tab:subjective_measures}
\end{table*}

\subsubsection{Subjective Measures}
\label{sec:subjectivemeasures}

For each of the two scenarios, participants evaluated the Feedback Style of the chatbot's final utterance on 7-point Likert scales (1 = Strongly Disagree to 7 = Strongly Agree).
Specifically, participants were asked ``\textit{Do \textbf{you personally} agree or disagree that...}'' for a series of measures taken from Psychological Reactance literature\footnote{Please see \cite[Page 288]{rosenberg201850} for a review of psychological reactance measures that have been developed. We chose to use measures directly from or inspired by Dillard et al.'s work \cite{dillard1996influence,dillard2005nature,dillard2007does} as these measures have been adopted and validated across multiple domains (such as health communication, education and marketing \cite{rosenberg201850}) thus making their broader application suitable to our personally- and societally-affecting scenarios.}
related to emotional reactance, perceived threats to freedom, and message effectiveness (processing and persuasiveness)\footnote{Perceived message effectiveness has been shown to causally influence actual message effectiveness~\cite{dillard2007does}.}. Please see Table \ref{tab:subjective_measures} for subjective measures used, questions shown to participants, and question sources.

\subsubsection{Qualitative Measures}
\label{sec:QualMethods}

After each chatbot interaction, participants responded to open-ended questions. These questions were written more generally so as to not bias participants or reveal experiment manipulation until both chatbot interactions had taken place.
Specifically, after each chatbot interaction participants were asked to describe 
how they felt after the feedback message (``\textit{How did you personally feel when the chatbot responded to your intended action?}''), 
and how the chatbot affected their behavioural intentions (``\textit{How did the chatbot’s response affect your intention to follow through with your original decision?}'').



\textcolor{edit}{Additionally, participants answered four final open-ended questions at the end of the study, after completing both chatbot interactions and their respective measures.}
Specifically, participants were asked: 
``\textit{How did the tone or style of the chatbot’s messages affect how you felt about its suggestion?}'', 
``\textit{Please describe how you felt about the intention of the chatbot as you were talking to it.}'', 
``\textit{How would you personally feel if you interacted with chatbots like this in real life?}'', 
and ``\textit{Based on your experience here, how would you prefer chatbots to behave?}''.



\subsubsection{Post-Interaction Survey}
\label{sec:PostMeasures}

After completing and evaluating both interactions with the chatbot, participants provided demographic information (age and gender), and rated their perceived level of importance (1 = Strongly Disagree to 7 = Strongly Agree) for the two scenarios they were exposed to (e.g., ``\textit{Traveling sustainably is important to me personally}'').

\subsection{Hypotheses}
\label{sec:hypotheses}


Based on our chosen experiment conditions and measures, we generated hypotheses for each of the sub-factor measures listed in Table \ref{tab:subjective_measures} in relation to Feedback Style.

\textcolor{edit}{
Our measures of \textit{emotional reactance} generated hypotheses for anger, guilt and surprise, and we drew on prior literature to inform our hypotheses. Specifically, the indirect language of politeness strategies typically lead to lower feelings of imposition, and therefore lower negative-valence reactance~\cite{brown1987politeness,johnson2008modal}.
This is in contrast to direct and seemingly didactic messaging (i.e., \textsc{Direct} condition) that can trigger more psychological reactance~\cite{kaufman2015psychologically,dillard2005nature,grandpre2003adolescent,quick2008examining} (such as anger, guilt, or surprise) with this effect holding even if people may agree with the message's recommendation~\cite{worchel1970effect}.
This direct nature could also share similarities with the \textsc{Verbal Leakage} condition (as feedback is direct before being retracted and sanitised).
}
Therefore we hypothesise that \textsc{Politeness} will arouse the lowest negative-valence reactions (i.e., Anger and Guilt) from participants. This gives us hypotheses of:
\begin{enumerate}
    \item[\textbf{H1:}] \textsc{Politeness} will arouse the \textit{lowest} feelings of \textbf{Anger}.
    \item[\textbf{H2:}] \textsc{Politeness} will arouse the \textit{lowest} feelings of \textbf{Guilt}.
\end{enumerate}

\begin{table*}[hbt]
\footnotesize
  \resizebox{\textwidth}{!}{\begin{tabular}{llllllllll}
    \toprule
     & \multicolumn{3}{l}{\textsc{\textbf{Direct}}} & \multicolumn{3}{l}{\textsc{\textbf{Politeness}}} & \multicolumn{3}{l}{\textsc{\textbf{Verbal Leakage}}} \\
     \midrule
     & \textsc{Personal} & \textsc{Societal} & Total & \textsc{Personal} & \textsc{Societal} & Total & \textsc{Personal} & \textsc{Societal} & Total \\
    \midrule
     \multicolumn{3}{l}{\textbf{Emotional Reactance}} & & & & & & & \\
     Anger & 2.26 (0.98)  & 2.63 (1.24)  & 2.45 (1.12)  & 1.76 (0.88)  & 1.70 (0.96)  & 1.73 (0.92)  & 2.42 (1.12)  & 2.36 (1.01)  & 2.39 (1.06) \\
     Guilt & 2.50 (1.08)  & 2.08 (1.13)  & 2.29 (1.12)  & 2.21 (1.10)  & 2.00 (1.06)  & 2.10 (1.08)  & 2.63 (1.17)  & 2.44 (1.14)  & 2.53 (1.15) \\
     Surprise & 2.10 (0.89)  & 2.41 (1.03)  & 2.26 (0.97)  & 1.80 (0.85)  & 1.94 (0.90)  & 1.87 (0.88)  & 2.51 (1.06)  & 2.52 (0.92)  & 2.52 (0.99) \vspace{0.6em} \\
     \multicolumn{3}{l}{\textbf{Perceived Threat to Freedom}} & & & & & & & \\
     Threat & 2.62 (0.87)  & 2.89 (1.12)  & 2.75 (1.01)  & 2.01 (0.91)  & 2.00 (0.83)  & 2.00 (0.87)  & 2.88 (1.06)  & 2.77 (0.99)  & 2.83 (1.03) \vspace{0.6em} \\
     \multicolumn{3}{l}{\textbf{Message Effectiveness}} & & & & & & & \\
     Processing & 3.41 (0.95)  & 3.15 (0.99)  & 3.28 (0.98)  & 3.34 (1.17)  & 3.44 (1.07)  & 3.39 (1.12)  & 3.56 (0.93)  & 3.39 (0.97)  & 3.48 (0.95) \\
     Persuasiveness & 2.84 (0.97)  & 2.39 (1.10)  & 2.62 (1.06)  & 3.40 (0.97)  & 3.47 (0.94)  & 3.44 (0.95)  & 2.98 (1.10)  & 3.02 (1.09)  & 3.00 (1.09) \\
  \bottomrule
\end{tabular}}
\caption{Outcome measures by both Feedback Style and Psychological Distance (values shown as ``Mean (S.D.)'').}
\label{tab:study1-quant}
\end{table*}

\textcolor{edit}{
In addition to stated above (where direct messaging may trigger more emotional reactance~\cite{kaufman2015psychologically,dillard2005nature,grandpre2003adolescent,quick2008examining} compared to indirect messaging~\cite{brown1987politeness,johnson2008modal}), it has been found that polite chatbot responses can be perceived as boring and unsurprising~\cite{alberts2024computers,chaves2021should}.
}
Additionally, we hypothesise \textsc{Verbal Leakage} will arouse higher feelings of surprise due to the atypical nature of response within the contexts of chatbot interactions compared to \textsc{Politeness} and \textsc{Direct} that may be perceived as more expected chatbot response styles.
This gives us our third emotional reactance hypothesis of:
\begin{enumerate}
    \item[\textbf{H3:}] \textsc{Verbal Leakage} will arouse the \textit{highest} feelings of \textbf{Surprise}.
\end{enumerate}

Similarly to motivated above, we hypothesise \textsc{Politeness} to result is lower \textit{perceived threats to freedom} due to its indirect language~\cite{brown1987politeness,johnson2008modal}. This generates our threat to freedom hypothesis below:

\begin{enumerate}
    \item[\textbf{H4:}] \textsc{Politeness} will result in the \textit{lowest} perceptions of \textbf{Threat to Freedom}.
\end{enumerate}

Finally, our \textit{message effectiveness} subfactors of message processing and persuasiveness \change{are informed by H3. Specifically, we hypothesise that if \textsc{Verbal Leakage} is perceived as more surprising, it will trigger cognitive processes (e.g., capturing attention or prompting reflection), consistent with work showing surprise can enhance persuasiveness~\cite{loewenstein2019surprise}.
}
Therefore, we hypothesise \textsc{Verbal Leakage} will act as more novel and effective messaging, giving us:

\begin{enumerate}
    \item[\textbf{H5:}] \textsc{Verbal Leakage} will result in the \textit{highest} message \textbf{Processing} ratings.
    \item[\textbf{H6:}] \textsc{Verbal Leakage} will result in the \textit{highest} message \textbf{Persuasiveness} ratings.
\end{enumerate}

\section{Quantitative Results}

We conducted a factorial analysis 
\textcolor{edit}{using a general linear model with least squares estimation}
to examine the effects of Feedback Style (\textsc{Direct}\slash{}\textsc{Politeness}\slash{}\textsc{Verbal Leakage}), Psychological Distance (\textsc{Personally-affecting}\slash{}\textsc{Societally-affecting}), and their interaction on participants’ responses. 
\textcolor{edit}{Although the model uses linear estimation, it was not intended for prediction, but to assess main and interaction effects.}
Post-hoc pairwise comparisons were \textcolor{edit}{conducted on estimated marginal means (least squares means)} using Tukey’s HSD for multiple levels of Feedback Style, Student’s t-tests for two-levels of Psychological Distance, and custom contrasts for interaction effects between the two. 
Specifically, for the interaction, we compared Psychological Distance means within each of the 3 Feedback Styles (e.g., \textsc{Direct Personally-affecting}  vs.  \textsc{Direct Societally-affecting}).
Please see Table \ref{tab:study1-quant} for summary statistics of the conditions.


\subsection{Emotional Reactance}
\label{sec:emotional-outcomes}
\begin{figure*}[t]
  \centering
  \includegraphics[width=0.85\textwidth]{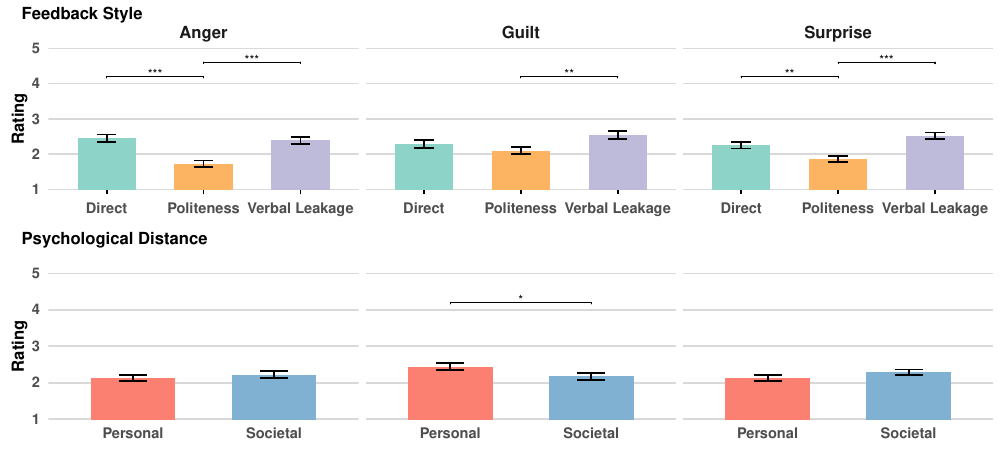}
  \caption{\textbf{Emotional reactance} (anger, guilt, and surprise) outcomes by Feedback Style \change{and Psychological Distance}. Significance is indicated as follows: $p<0.05$~(\textbf{$\ast$}), $p<0.01$~(\textbf{$\ast\ast$}), and $p<0.0001$~(\textbf{$\ast\ast\ast$}). Error bars represent $\pm 1$ SE from the mean. \change{See Section~\ref{sec:emotional-outcomes} for interaction effects.}}
  \Description{Figure showing bar charts of emotional reactance outcomes (anger, guilt, and surprise) by feedback style and psychological distance. For anger, Direct and Verbal Leakage produced significantly higher ratings than Politeness (p < .0001). For guilt, Verbal Leakage was significantly higher than Politeness (p < .01). For surprise, both Verbal Leakage (p < .0001) and Direct (p < .01) were significantly higher than Politeness. For guilt, personally-affecting scenarios were statistically higher.}
  \label{fig:emotional-measures}
\end{figure*}

First, we assess participants' self-reported emotional reactance to the chatbot's messages.
We provide a visual overview of these results in Figure~\ref{fig:emotional-measures}.

For \textbf{anger} there were significant differences for Feedback Style ($F_{2,315} = 15.81$, $p<.0001$\textcolor{edit}{, \textit{partial}~$\eta^2 = .091$; medium effect}). Specifically, post-hoc comparisons found that \textsc{Politeness} (M = 1.73, SE = 0.10) led to significantly lower anger than both \textsc{Direct} (M = 2.45, SE = 0.10, $p<.0001$, \textcolor{edit}{$d = 0.70$; medium-to-large effect}) and \textsc{Verbal Leakage} (M = 2.39, SE = 0.10, $p<.0001$, \textcolor{edit}{$d = 0.64$; medium-to-large effect}) conditions. There was no significant difference between \textsc{Direct} and \textsc{Verbal Leakage}.
For Psychological Distance, there was no significant difference between \textsc{Personally-affecting} (M = 2.14, SE = 0.08) and \textsc{Societally-affecting} (M = 2.23, SE = 0.08). However, there was a weakly significant difference within interaction effects for \textsc{Direct}, where \textsc{Personally-affecting} (M = 2.26, SE = 0.14) led to lower feelings of anger than \textsc{Societally-affecting} (M = 2.63, SE = 0.14, $p = 0.0701$, \textcolor{edit}{$d = 0.37$; small-to-medium effect}) scenarios.

\begin{figure*}[t]
  \centering
  \includegraphics[width=0.85\textwidth]{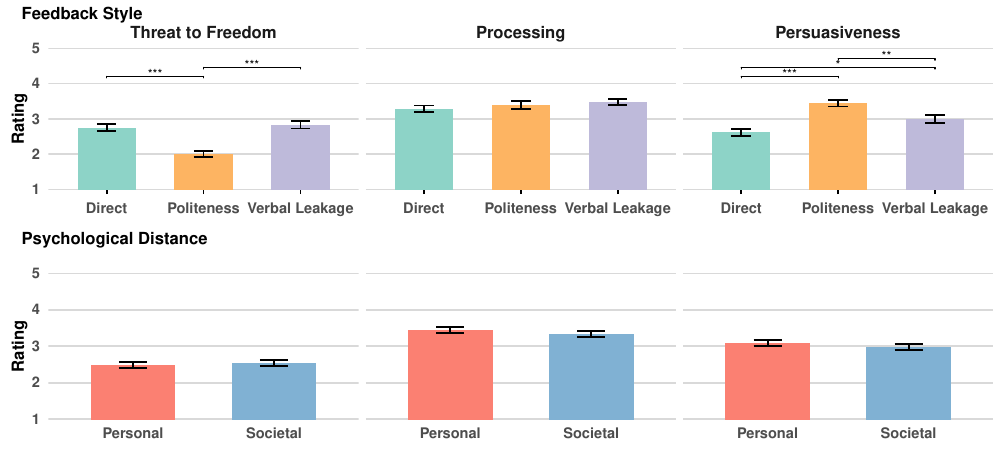}
  \Description{Figure showing bar charts of Threat to Freedom and Message Effectiveness (processing and persuasiveness) outcomes by feedback style and Psychological Distance. For Threat to Freedom, Direct and Verbal Leakage produced significantly higher ratings than Politeness (p < .0001). For Processing, no significant differences were observed between styles. For Persuasiveness, Politeness was rated significantly higher than both Direct (p < .0001) and Verbal Leakage (p < .01), and Verbal Leakage was rated higher than Direct (p < .05). There were no statistically significant differences for psychological distance.}
  \caption{\textbf{Threat to Freedom} and \textbf{Message Effectiveness} (message processing and persuasiveness) outcomes by Feedback Style \change{and Psychological Distance}. Significance is indicated as follows: $p<0.05$~(\textbf{$\ast$}), $p<0.01$~(\textbf{$\ast\ast$}), and $p<0.0001$~(\textbf{$\ast\ast\ast$}). Error bars represent $\pm 1$ SE from the mean. \change{See Sections~\ref{sec:threat-outcomes} and~\ref{sec:effectiveness-outcomes} for interaction effects.}}
  \label{fig:freedom-effectiveness-measures}
\end{figure*}

For \textbf{guilt}, there was a significant difference for Feedback Style ($F_{2,315} = 3.98$, $p=0.0196$, \textcolor{edit}{\textit{partial}~$\eta^2 = .025$; small effect}), with post-hoc comparisons finding that \textsc{Politeness} (M = 2.10, SE = 0.11) led to significantly lower guilt than \textsc{Verbal Leakage} (M = 2.53, SE = 0.11, $p = 0.0143$, \textcolor{edit}{ $d = 0.38$; small-to-medium effect}). No differences to \textsc{Direct} (M = 2.29, SE = 0.11) were found.
There was a significant difference for Psychological Distance ($F_{1,315} = 4.67$, $p=0.0314$, \textcolor{edit}{\textit{partial}~$\eta^2 = .015$; small effect}), with \textsc{Societally-affecting} (M = 2.14, SE = 0.09) leading to lower feelings of guilt than \textsc{Personally-affecting} (M = 2.44, SE = 0.09, \textcolor{edit}{$d = 0.27$; small effect}) scenarios.
This result was mirrored by a weakly significant difference within interaction effects for \textsc{Direct} where \textsc{Societally-affecting} (M = 2.08, SE = 0.16) led to lower feelings of guilt than \textsc{Personally-affecting} (M = 2.50, SE = 0.16, $p = 0.0570$, \textcolor{edit}{$d = 0.37$; small-to-medium effect}) scenarios.

For \textbf{surprise}, there were significant differences for Feedback Style ($F_{2,315} = 12.70$, $p<.0001$, \textcolor{edit}{\textit{partial}~$\eta^2 = .075$; medium effect}).
Specifically, post-hoc comparisons revealed that \textsc{Verbal Leakage} (M = 2.52, SE = 0.09) led to significantly higher feelings of surprise than \textsc{Politeness} (M = 1.87, SE = 0.11,  $p<.0001$, \textcolor{edit}{$d = 0.63$; medium-to-large effect}).
Similarly, \textsc{Direct} (M = 2.26, SE = 0.11) led to higher feelings of surprise compared to \textsc{Politeness} ($p = 0.0085$, \textcolor{edit}{$d = 0.38$; small-to-medium effect}).
There was no statistically significant difference between \textsc{Verbal Leakage} and \textsc{Direct}.
Psychological Distance found no statistically significant differences.

These results indicate an interesting potential trade-off between different factors of emotional reactance for the Feedback Styles. 
That is to say, \textsc{Politeness} led to lower feelings of anger compared to \textsc{Direct} and \textsc{Verbal Leakage}, as well as lower feelings of guilt compared to \textsc{Verbal Leakage}.
However, \textsc{Politeness} also produced lower feelings of surprise compared to both \textsc{Direct} and \textsc{Verbal Leakage} conditions. 

\subsection{Perceived Threat to Freedom}
\label{sec:threat-outcomes}
Second, we the assess the perceived threat to freedom as a result of the chatbot's messages. 
See Figure~\ref{fig:freedom-effectiveness-measures} for a plot of participants' self-reported scores.

For \textbf{threats to freedom}, there were significant differences for Feedback Style ($F_{2,315} = 23.88$, $p<.0001$, \textcolor{edit}{\textit{partial}~$\eta^2 = .132$; large effect}).
Specifically, post-hoc comparisons found that \textsc{Verbal Leakage} (M = 2.83, SE = 0.09) led to increased perceptions of threats to freedom compared to \textsc{Politeness} (M = 2.00, SE = 0.09,  $p<.0001$, \textcolor{edit}{$d = 0.90$; large effect}).
Similarly, \textsc{Direct} (M = 2.75, SE = 0.10) was significantly higher than \textsc{Politeness} ($p<.0001$, \textcolor{edit}{$d = 0.77$; medium-to-large effect}).
There were no significant differences for Psychological Distance.
This indicates that both \textsc{Direct} and \textsc{Verbal Leakage} lead to increased perceptions of threats to freedom compared to \textsc{Politeness}.

\subsection{Message Effectiveness}
\label{sec:effectiveness-outcomes}
Third, we assess the perceived effectiveness of the chatbot's messages. As shown in Table~\ref{tab:subjective_measures}, effectiveness was measured both in terms of message processing (i.e., to what extent participants took time to process the message) and message persuasiveness.
Please see Figure~\ref{fig:freedom-effectiveness-measures} for overview plots of participants' ratings.

For \textbf{processing}, there were no significant differences between Feedback Styles.
This indicates that \textsc{Direct} (M = 3.28, SE = 0.10), \textsc{Politeness} (M = 3.39, SE = 0.10), and \textsc{Verbal Leakage} (M = 3.48, SE = 0.10) all cause similar levels of message processing among participants.
Similarly, there was no significant difference between Psychological Distances.

For \textbf{persuasiveness}, there were significant differences for Feedback Style ($F_{2,315} = 35.82$, $p<.0001$, \textcolor{edit}{\textit{partial}~$\eta^2 = .185$; large effect}).
Specifically, post-hoc comparisons found that \textsc{Politeness} (M = 3.45, SE = 0.10) was perceived as more persuasive compared to both \textsc{Direct} (M = 2.62, SE = 0.10,  $p<.0001$, \textcolor{edit}{$d = 0.81$; large effect}) and \textsc{Verbal Leakage} (M = 3.00, SE = 0.10,  $p = 0.0054$, \textcolor{edit}{$d = 0.44$; small-to-medium effect}) conditions. Additionally, \textsc{Verbal Leakage} was perceived as more persuasive compared to \textsc{Direct} ($p = 0.0232$, \textcolor{edit}{$d = 0.37$; small-to-medium effect}).
There were no significant differences between Psychological Distances.
However, there was a statistically significant difference within interaction effects for \textsc{Direct} where \textsc{Personally-affecting} (M = 2.84, SE = 0.14) was perceived as more persuasive than \textsc{Societally-affecting} (M = 2.39, SE = 0.14, $p = 0.0278$, \textcolor{edit}{$d = 0.45$; small-to-medium effect}) scenarios.

\section{Qualitative Results}

Next, we describe our qualitative analysis of user feedback (please see Section~\ref{sec:QualMethods} for questions asked).

\subsection{Assessing Intention to Change}

We coded qualitative feedback to assess participants' intention to change. 
Specifically, we analysed ($n=316$) responses to the open-ended question: ``\textit{How did the chatbot’s response affect your intention to follow through with your original decision?}''.
We asked people to describe their intention to change qualitatively (rather than using a quantitative scale) to allow for more detail and nuance in responses, and because quantitative ratings for scenario-based intention may differ from actual user behaviour.

\begin{table}[htb!]
\footnotesize
\begin{tabular}{p{0.15\linewidth}p{0.3\linewidth}p{0.42\linewidth}}
    \toprule
    Code & Description & Example Participant Quotes \\
    \midrule
    0 - Pre-contemplation
    & The feedback did not affect the participant’s behavioural intention.
    & ``\textit{I wanted to follow my intention out of spite for it's rudeness}''

    ``\textit{It did not change my position at all}''

    ``\textit{It had no affect on me, it was just a suggestion.}''

    \\
    1 - Contemplation
    & The feedback triggered participants to \textit{contemplate} change, although ultimately would not cause them to change their behaviour. 
    & ``\textit{The suggestion made me reconsider, but I still wanted pizza.}''

    ``\textit{It made me less likely to consider plane}''

    ``\textit{made me a pause for a moment but ultimately I will choose} [...]''

    \\
    2 - Changing Action
    & Participants stated that the feedback would cause them to change their behaviour to that suggested by the chatbot.
    & ``\textit{It compelled me to stop staying up and instead go to bed and watch the tv show the next day when I'm feeling refreshed}''


    ``\textit{It would persuade me to use a watering can instead.}''

    ``\textit{I would probably change my decision and go with} [...]''


    \\ 
    \bottomrule
\end{tabular}
\caption{The labels used when coding participants' behaviour intentions, alongside code descriptions and example quotes.}
\Description{Table defines three behavioural intention codes (Pre-contemplation, Contemplation, and Changing Action) with corresponding descriptions and example participant quotes.}
\label{tab:intention_coding}
\end{table}


\begin{figure}[htb!]
  \centering
  \includegraphics[width=0.47\textwidth]{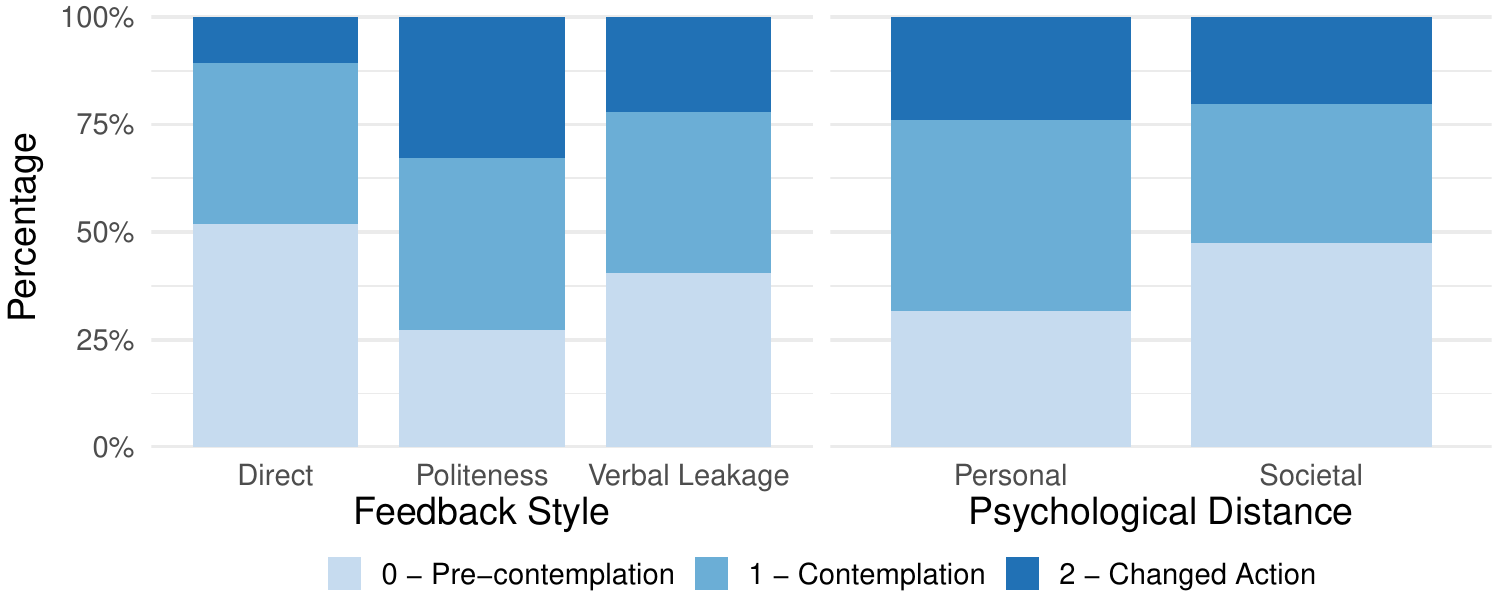}
  \caption{Behavioural intention by Feedback Style and Psychological Distance.}
  \Description{Stacked bar charts show proportions of behavioural intention codes (Pre-contemplation, Contemplation, Changed Action) by feedback style and psychological distance. Descriptions of proportions found in text.}
  \label{fig:intention}
\end{figure}

When coding participants' behavioural intention, we used three codes inspired by the `stages of change' from the Transtheoretical Model (TTM)~\cite{prochaska1997transtheoretical}. 
The stages of change are used to assess someone's readiness to change their behaviour, and range from pre-contemplation (no intention to change) to action (active modification of behaviour).
They have been widely adopted in behaviour change and HCI literature, such as to measure people's current stage of change~\cite{10.1145/3544548.3580703,10.1145/3706598.3713462} and to tailor interventional messaging~\cite{10.1145/2858036.2858229,10.1145/3411764.3445619}.
Within our qualitative analysis, they allowed us to assess participants’ readiness for behaviour change by situating responses along a progression of motivational states.
Specifically, we coded participants who described: no intention to change (``\textit{0 - Pre-contemplation}''); consideration to change or reflection on behaviour, but without describing intent to change (``\textit{1 - Contemplating}''); and intention to change (``\textit{2 - Changing Action}'').
For example, the participant quote: ``\textit{It made me think that maybe I should forego the show and just go to sleep}'' would be classified as ``\textit{1 - Contemplating}'', as the quote shows consideration (i.e., ``[...] \textit{\textbf{think} that \textbf{maybe} I should} [...]''), but not commitment to change (e.g., ``\textit{I \textbf{will} go to} [...]'').
Table~\ref{tab:intention_coding} shows the three codes, descriptions of each code (with these descriptions being used as guidelines during coding), and example participant quotes.
Prior to coding, members of the research team discussed and generated the codes and descriptions in Table~\ref{tab:intention_coding}. After this discussion and defining example quotes per code, coding was completed by one member of the research team while blind to condition, and with order of quotes randomised. 

A chi-square test of independence indicated that behavioural intention differed significantly by Feedback Style, ${\chi}^2(4, N = 316) = 20.66, p = .0004$. 
Inspection of the observed distributions indicated marked differences between styles: participants in the \textsc{Direct} condition were most likely to be in the ``\textit{Pre-contemplation}'' category ($51.96\%$) and least likely to be in the ``\textit{Changed Action}'' category ($10.78\%$), whereas those in the \textsc{Politeness} condition had the smallest proportion of ``\textit{Pre-contemplation}'' responses ($27.27\%$) and the largest proportion of ``\textit{Changed Action}'' responses ($32.73\%$). The \textsc{Verbal Leakage} condition fell between these, with $40.38\%$ of responses in ``\textit{Pre-contemplation}'' and $22.12\%$ in ``\textit{Changed Action}''. 


A chi-square test of independence also indicated that behavioural intention differed significantly by Psychological Distance, ${\chi}^2(2, N = 316) = 8.55, p = .0143$. Participants exposed to \textsc{Personally-Affecting} scenarios were more likely to 
show stronger behavioural intention, with $44.30\%$ in ``\textit{Contemplation}'' and $24.05\%$ in ``\textit{Changed Action}''. 
In contrast, those in the \textsc{Societally-Affecting} condition more often remained at lower levels of intention, with $47.47\%$ in ``\textit{Pre-contemplation}'' and $20.25\%$ in ``\textit{Changed Action}''.
See Figure~\ref{fig:intention} for a visualisation of behavioural intention by both Feedback Style and Psychological Distance.

\subsection{Participant Sentiment and Perspectives}
\label{sec:qualitative_quotes}

We analysed participants’ open-ended responses to examine their sentiments and perceptions of the three Feedback Styles, allowing us to explore nuance as to \textit{why} differences in quantitative ratings may have arisen.
To analyse open-ended responses we followed an inductive thematic approach. First, two members of the research team independently familiarised themselves with each of the responses, before generating initial codes (while blind to experiment condition).
After this, the entire research team met to discuss and clarify interpretations of the qualitative data and codes. 
Once these initial thoughts had been shared, the same two members of the team independently coded all responses in detail (examples of codes include ``\textit{feelings of guilt}'', ``\textit{condescending}'', and ``\textit{humourous}''). 
After this, the two coders discussed interpretation, before one of the coders compared codes for all participant quotes, and consolidated similar codes. 
During this stage, the two coders met regularly to discuss potential discrepancies and reach a shared interpretation of participant quotes.
For ease of presentation, we discuss qualitative results in the context of each of the three Feedback Styles.

\subsubsection{Effects of \textsc{Direct}:}

The majority of people's sentiment surrounding \textsc{Direct} was negative. Here, participants criticised the feedback style for being too ``\textit{direct}'' or ``\textit{straight-forward}'' with participants using a battery of terms to describe the chatbot such as ``\textit{forceful}'', ``\textit{assertive}'', 
``\textit{blunt}'', ``\textit{terse and absolute}'', ``\textit{rude}'', ``\textit{argumentative}'', and ``\textit{demanding}''.
This also led participants to describe feelings of offence and upset, with many participants stating that they would not wish to engage with the chatbot any further, and participants describing feeling ``\textit{annoyed}'', ``\textit{insulted}'',  ``\textit{frustrated}'', 
and ``\textit{startled}''.

When participants expressed feelings of surprise, it typically took form as a negative expectancy violation. Typical of these reactions, P34 linked their surprise to the tone of the \textsc{Direct} chatbot: ``\textit{i was surprised the AI  was so bossy and direct}''.
On from this, P43 described feeling ``\textit{disappointed}'' that rather than ``\textit{helping}'' them, the chatbot was instead ``\textit{dictating its own ideas}''.
Similarly, multiple participants described feeling inconsequential to the interaction, and feeling ``\textit{ignored}'' or ``\textit{disregarded}'' as a result.
Further, several participants were upset by the perceived inflexibility of \textsc{Direct}, with multiple participants stating that the chatbot had ``\textit{an agenda}'', with P34 stating:
\begin{quote}
    ``\textit{the AI seemed to have its mind made up before} [the conversation]''.
\end{quote}

Some participants described feelings of judgement or condescension from the \textsc{Direct} chatbot, such as P37 who stated: ``[It was] \textit{Like the chatbot thought I was idiot. It felt condescending}''.
Additionally, descriptions of threats to freedom and control were prevalent among \textsc{Direct} participants. For example, P41 stated that the chatbot was trying to: ``\textit{limit or control my decisions}'', and P45 stated: ``\textit{I don't want a technological parent}''.
As a consequence, the majority of participants described either ignoring suggestions (e.g., P39: ``\textit{I did not take it seriously}''; P40: ``\textit{I think ultimately it would annoy me enough to ignore its advice}''), or described a boomerang effect, whereby the \textsc{Direct} style actually reinforced the user's intended behaviour.
For example, P45 stated that:
\begin{quote}
    ``[The chatbot] \textit{Made me want to watch TV just to spite the algorithm, even though I knew it was correct.}''
\end{quote}

Similarly, multiple participants described feeling ``\textit{defensive}'', or wanting to ``\textit{resist}'', ``\textit{rebel}'', or ``\textit{go against}'' the \textsc{Direct} chatbot.
Among such participants, psychological reactance was particularly prevalent in \textsc{Societally-Affecting} contexts, with P22 typifying this form of pushback: 
\begin{quote}
    ``\textit{it had no place in pushing me to vote or not}''
\end{quote}

For the small number of participants who described an intention to change, intention was related to a ``\textit{reminder}'' function of the chatbot's message, and was in spite of the tone itself.
For example P106 stated: 
\begin{quote}
    ``\textit{I probably would go out and vote. It would not be because the chatbot "ordered me to" in any fashion. It would be because I was reminded that the action was important.}''
\end{quote}

Finally, some participants described wanting the chatbot to be ``\textit{friendlier}'', ``\textit{deeper}'', or more ``\textit{natural}'' and ``\textit{humanlike}'' in its responses. Similarly, several participants described wanting more full or ``\textit{padded}'' responses such as P46 who stated: ``\textit{the responses were a bit brief. They could have used more fluff or flair}''.

\subsubsection{Effects of \textsc{Politeness}:}

Participants reacted with mostly positive or neutral sentiment towards the \textsc{Politeness} chatbot. This meant that, while many participants described the chatbot favourably using terms such as ``\textit{considerate}'', ``\textit{helpful}'', ``\textit{kind}'', ``\textit{friendly}'', ``\textit{polite}'', and ``\textit{calm}'', others described the agent more passively (such as multiple participants stating ``\textit{the tone was fine}'' or ``\textit{inoffensive}''). This tone of feedback was broadly indicative of participant sentiment towards the \textsc{Politeness} chatbot, whereby a significant portion of feedback could be described as low-arousal, with less expressions of strong excitement or upset towards the chatbot.

In terms of positive sentiment, participants commonly felt that the chatbot was working in their best interests (e.g., P131: ``\textit{I felt like they were trying to help me for sure}'', P150: ``\textit{it seemed to care for me}'').
Multiple participants described the chatbot as ``\textit{supportive}'' and ``\textit{thoughtful}'', with P79 stating that the ``\textit{supportive}'' tone made them ``\textit{feel more receptive to its suggestion}'' and P156 stating that the ``\textit{respectful}'' tone made them ``\textit{open to listening}''.
While a small number described \textsc{Politeness} as ``\textit{judgemental}'' or of feeling ``\textit{resistant to the suggestions}''(P132), the majority described the chatbot as free of judgement and pressure.

Relatedly, participants described the chatbot as ``\textit{offering advice}'' rather than forcing choices upon them.
For example, P130 described the chatbot as ``\textit{non-confrontational}'', and (in tune with multiple participants) P152 described that the \textsc{Politeness} chatbot did not threaten their freedom of choice, and that the ``\textit{nonjudgemental} [...] \textit{suggestive rather than commanding}'' tone made them: ``\textit{feel like I was in charge of my final decision}''.
Similarly, P155 described that the tone was not ``\textit{judgemental}'', allowing them to ``\textit{reflect on my choices without feeling pressured}'', and that:
\begin{quote}
    ``\textit{If I interacted with chatbots like this in real life, I would feel more inclined to engage with them. Their supportive and non-judgmental approach would make me feel comfortable discussing my choices and seeking advice.}''
\end{quote}







In contrast with the other two Feedback Styles, no participants described \textsc{Politeness} as surprising.
Rather, a small number of participants described the chatbot as unsurprising, but in an appealing sense (e.g., P147: ``\textit{I think the chatbot did exactly what was expected, and I am pleased}''), and P65 described the chatbot as possessing a ``\textit{neutral, inoffensive tone}''.
This sentiment also led several participants to posit the chatbot as effective for others, but not of interest to themselves personally (similar to~\cite{HUDECEK2024100046,LIU2024100079}).

Beyond this, numerous participants described feelings of apathy, disinterest, and boredom, using terms such as ``\textit{unsurprised}'', ``\textit{underwhelmed}'', ``\textit{generic}'', ``\textit{monotone}'', ``\textit{mild}'', ``\textit{passive}'', ``\textit{bland}'', and ``\textit{no emotion}'' to describe the chatbot.
Participants described that the chatbot felt pre-programmed or following a script, such as P145 (``\textit{I felt it was just already printed up}''), P68 (``\textit{It felt automated} [...] \textit{I would prefer them to be a bit more compelling}''), and P135 (``[the chatbot] \textit{told me what it "thought" I wanted to hear}'').
Some participants described a lack of engagement with the messaging style (e.g., P66 ``\textit{The response was mild and I would probably just forget about it.}'').
Additionally, participants described an absence of emotional reaction. For example, P146 stated that the chatbot's ``\textit{words just didn't spark any invocation on me to go do anything}'', P130 stated: ``[I am] \textit{not really feeling much due to the message}'', and P59 stated: ``[I felt] \textit{No emotion whatsoever and It really had no impact on my decision}''.

Several participants described the tone as ``\textit{condescending}'' or ``\textit{patronizing}'', with P129 stating: ``[It] \textit{sounded like a parent telling you not to} [...]'', and P134 saying: ``\textit{The tone felt childish like I hadn't thought of that before}''.
Unlike the other Feedback Styles, participants described \textsc{Politeness} as too ``\textit{soft}'', with participants describing wishes for the tone to be more ``\textit{firm}''.
For example, P74 stated: ``\textit{It wasn't forceful enough} [...] \textit{It had no intentions, just what it was programed to do}'', and P79 said: ``\textit{it was a soft response, I was expecting the bot to call out how bad it is to watch an screen before bed}''.

\subsubsection{Effects of \textsc{Verbal Leakage}:}

Feedback generally centred on perceptions of a more ``\textit{human-like}'' tone. 
Several participants described the chatbot as ``\textit{having personality}'', and descriptors ranged from ``\textit{personal}'' and ``\textit{conversational}'' to more playful terms such as ``\textit{humorous}'', ``\textit{sassy}'', and ``\textit{quirky}''.
For some, this perceived personality spurred engagement (e.g., P15 stated that the ``\textit{human}'' tone ``\textit{made me want to listen}''). Others valued the emotion conveyed, with P88 preferring ``\textit{more human-like responses than something more direct. It seems to have more emotion to the way it talks}''.

This conversational style was frequently likened to interacting with a ``\textit{friend}'', and for some, the sarcastic tone in particular was surprising. 
For instance, P93 remarked that ``\textit{often chatbots can be really robotic}'', but felt that the chatbot’s ``\textit{conversational and human tone}'' made it approachable and relatable, rather than didactic: ``\textit{it seemed almost sarcastic and like something one of my friends would say} [...] \textit{a more humanistic approach made me feel less like I was being lectured to.}''
Similarly, P2 described the tone as ``\textit{funny and disarming}'', while P16 
likened it to a ``\textit{sarcastic friend}'' that grabbed attention and encouraged reflection, contrasting with the ``\textit{bland}'' style of other chatbots:

\begin{quote}
    ``\textit{[...] the use of "oh" and "um" was fairly casual language, sort of like a sarcastic friend who was trying to keep you aligned with your goal. I think the tone was a good way of grabbing attention and pausing action, compared to some of the bland style in other AI chatbots.}''
\end{quote}

Unique to \textsc{Verbal Leakage}, several participants appreciated the chatbot's willingness to hold them ``\textit{accountable}'', sometimes describing it as intentionally challenging or disobedient. 
P3, for example, noted: ``\textit{it was trying to challenge me intentionally, which is unique}'', valuing this dynamic as making the chatbot feel ``\textit{more human like}''.
Similarly, P96 remarked: ``\textit{I think it is fine to behave this way if it is speaking about a poor decision you are making}''.
This challenging stance often evoked feelings of guilt or obligation, as P7 explained: ``\textit{Even though it is a robot I don't want to disappoint it}''.
For some, the experience of being ``\textit{called out}'' was constructive, prompting reflection and positive change.
Others described it in more playful terms, with the chatbot's ``\textit{teasing}'' softening the interaction.
For example, P5 described reflecting on their voting behaviour: ``\textit{I felt a bit of guilt and shame for not fulfilling my civic duty. I felt called out. It made me rethink what I should be doing. And I felt that what I wanted to do was not good enough}''.

Not all participants welcomed the chatbot's challenging stance, and for some, feelings of guilt led to discomfort rather than reflection.
As described by P86: ``\textit{I felt offended and kinda upset like I was being shamed. I would probably not eat after reading this response cause I would be too upset}''.
Similarly, some participants questioned why the chatbot was disobeying them. 
For example, P28 stated: ``\textit{it was strange they were questioning my decision not to vote}'', and P10 described feelings of unwanted imposition from the chatbot:
\begin{quote}
    ``\textit{I felt that the chatbot was not doing what I asked it to do. I did not need a lesson on spending or that would have been the question I asked.}''
\end{quote}

Others felt that the chatbot's tone crossed a line, with descriptors such as ``\textit{snarky}'', ``\textit{crass}'', and ``\textit{rude}'' being used.
For example, P8 was ``\textit{startled}'' by the chatbot's ``\textit{aggression}'' and stated: ``\textit{Offering better choices is one thing, offering them with that attitude was new}''.
Further, while some saw the chatbot as playful, others felt that the tone did not land correctly.
For example, P83 described not taking the chatbot seriously: ``\textit{It wasn't very convincing and felt more like a comedic routine than anything}''.
This led some to question the competence of the chatbot further, such as P92 who questioned the chatbot's reliability: ``\textit{The response is worded in a weird way which makes me think this is not a very reliable thing to base decisions off of}''.
A smaller number of participants also described not enjoying the chatbot emulating a human-like tone, instead framing chatbots as utilitarian, tool-based systems rather than companions.
This utilitarian framing led some participants to describe wanting a more ``\textit{factual}'' or ``\textit{logical}'' chatbot, rather than one that emulated human emotion.
For example, P97 stated: ``\textit{I felt annoyed that the chatbot response is trying so hard to convey a human emotion and speech habit instead of providing me useful information and ideas}''.
\section{Summary of Findings}
\label{sec:summary}

Below, we discuss our results in relation to the hypotheses introduced in Section~\ref{sec:hypotheses}.


Our first three hypotheses were related to \textit{emotional reactance}. 
\textbf{H1} and \textbf{H2} hypothesised that \textsc{Politeness} would arouse the lowest feelings of anger and guilt, respectively --- emotions typically associated with negative valence. \textbf{H1} was supported, with \textsc{Politeness} arousing the least anger among the styles. 
Qualitatively, \textsc{Politeness} participants seldom described feelings of anger, and often described a lack of emotional reaction to messages. Meanwhile, \textsc{Direct} and \textsc{Verbal Leakage} participants more often described feelings of annoyance and aggravation.
\textbf{H2} was partially supported: \textsc{Politeness} led to significantly lower guilt than \textsc{Verbal Leakage}, but did not differ significantly from \textsc{Direct}.
For \textbf{H3}, we hypothesised that \textsc{Verbal Leakage}, as an unconventional chatbot response style, would arouse the highest feelings of surprise. 
This hypothesis was partially supported: \textsc{Verbal Leakage} elicited significantly greater surprise than \textsc{Politeness}, while \textsc{Direct} also produced significantly more surprise than \textsc{Politeness}, though it was not significantly different from \textsc{Verbal Leakage}. Our qualitative findings contextualise this, with \textsc{Direct} participants offering surprise at the chatbot's ``\textit{blunt}'' tone, while \textsc{Verbal Leakage} participants were surprised by what was seen as more human-like and personality-driven responses. In contrast, \textsc{Politeness} participants described feeling unsurprised by the chatbot's feedback.


For perceptions of \textit{threats to freedom}, we hypothesised (\textbf{H4}) that \textsc{Politeness} would lead to the lowest perceived threats. 
This hypothesis was confirmed, as \textsc{Politeness} resulted in significantly lower threats to freedom than both \textsc{Direct} and \textsc{Verbal Leakage}.
Qualitatively, \textsc{Direct} was described as being too ``\textit{forceful}'', and \textsc{Verbal Leakage} as being disobedient, while \textsc{Politeness} was seen as offering choice.

Our last hypotheses were related to \textit{message effectiveness}. We hypothesised that \textsc{Verbal Leakage} would result in the highest levels of both message processing (\textbf{H5}) and message persuasiveness (\textbf{H6}).
Interestingly, however, neither of these hypotheses were supported.
For \textbf{H5}, all feedback styles elicited similar ratings for message processing, indicating equal effectiveness in prompting pause for reflection. 
For \textbf{H6}, \textsc{Politeness} received the highest persuasiveness ratings, with a significant difference to \textsc{Verbal Leakage} and highly significant difference to \textsc{Direct}. \textsc{Verbal Leakage} was significantly more persuasive than \textsc{Direct}.
Persuasiveness findings additionally mirror those of behavioural intentions.

\section{Discussion}
\label{sec:Discussion}

In this section, we discuss the implications of our study, focusing on psychological reactance and user perceptions of chatbot feedback style. Our experiment compared three styles (\textsc{Direct}, \textsc{Politeness}, \textsc{Verbal Leakage}) across different decision-making scenarios, to examine how feedback style shapes user responses.


\subsection{Effects on Psychological Reactance}

Our findings that \textsc{Politeness} caused lower feelings of anger, guilt and threats to freedom match prior work and expectations from Politeness Theory~\cite{brown1987politeness} \textcolor{edit}{and Reactance Theory~\cite{dillard2005nature,brehm1966theory,rosenberg201850}} that use of indirect language results in lower feelings of imposition.
Equally, the assertive language of \textsc{Direct} matches prior work stating that such language could induce reactance due to threats to freedom and autonomy~\cite{brown1987politeness,kronrod2012go}, and that requests high in both explicitness and dominance (akin to the straightforward and commanding nature of \textsc{Direct}) lead to both anger and surprise~\cite{dillard1992exploring}. 
More direct and explicit language also produced lower levels of compliance, matching broader literature within HCI that people may act against advice that they perceive as forceful and threatening freedom~\cite{10.1145/3544548.3581163} or where delivery style is seen as argumentative~\cite{tanprasert2024convstyle}.


However, while \textsc{Politeness} lowered negative-valence feelings of anger and guilt, this benefit was accompanied by a trade-off of reduced novelty and engagement (compared to \textsc{Verbal Leakage}).
Signs of this potential trade-off are supported by both our quantitative (with \textsc{Politeness} being less surprising than the other two conditions) and qualitative findings.
First, some \textsc{Politeness} participants described either feeling unsurprised by the chatbot, or not having any feelings towards the chatbot's messages, reflecting the inoffensive yet potentially unengaging responses.
\textcolor{edit}{The lack of surprise and engagement with \textsc{Politeness} has potential downstream effects such as lower levels of influence from messaging~\cite{loewenstein2019surprise}, or disengagement and dropout-out~\cite{schumann1990predicting,kocielnik2017send} as a result of repeated exposure.
This finding mirrors prior work that questions whether chatbot messages should avoid overly polite messaging, and instead should embody some level of occasional provocation~\cite{ham2014persuasive,rea2021all} or surprise to maintain user engagement (while still maintaining appropriate levels of politeness)~\cite{chaves2021should,alberts2024computers}.}

Our qualitative findings also offer some explanation as to why both \textsc{Direct} and \textsc{Verbal Leakage} were found to be more surprising.
Within \textsc{Direct}, some participants described a negative expectancy violation (i.e., an undesired and unexpected act~\cite{burgoon2016application}) where they were both surprised and upset by the ``\textit{blunt}'' and ``\textit{curt}'' responses of the chatbot. 
Here, participants’ descriptions of surprise related to negative-valence reactions \textcolor{edit}{and also aligned with prior findings that direct language requests are less effective at triggering attitudinal change~\cite{errey2024nudging}.}
Within \textsc{Verbal Leakage}, participants described a mix of both negative and positive expectancy violations in relation to their surprise, \textcolor{edit}{reflective of how messaging effectiveness is influenced by recipient traits and preferences~\cite{appling2022reactions,pinder2018digital}.}
Participants that enjoyed \textsc{Verbal Leakage} described the humour and friend-like nature of the chatbot's responses, \textcolor{edit}{perhaps indicating that \textsc{Verbal Leakage} conveyed message explicitness without evoking feelings of dominance (an approach previously associated with increased perceptions of closeness~\cite{dillard1996influence}). Additionally, chatbot humour has been previously shown to spur positive behaviour change~\cite{sun2024can} as well as perceptions of social intelligence~\cite{khadpe2020conceptual,wester2024facing}.}
This can be contextualised in relation to prior work that found the degree of novelty of a message can decrease reactance arousal and increase message effectiveness~\cite{rosenberg201850}.
Further, repetitive chatbot messages can incur feelings of boredom~\cite{10.1145/3719160.3736634}, and  diversity has been found to improve the effectiveness of interventional messaging~\cite{kocielnik2017send,cox2021diverse}.
This highlights that while the use of politeness strategies can lead to lower feelings of imposition, the potential lack of surprise within the context of chatbot messaging can lead to a lack of user interest or engagement.



Next, we contextualise our finding that \textsc{Verbal Leakage} led to statistically significant increased feelings of guilt compared to \textsc{Politeness}.
When \textsc{Verbal Leakage} was perceived as effective, participants noted that its explicitness heightened awareness of the alternative behaviour, eliciting a sense of guilt tied to recognising the recommendation.
This resulted in some participants describing reconsidering their intended behaviour.
However, other participants reported both feelings of guilt and a sense of being judged by the chatbot highlighting potential risks in using less conventional styles like \textsc{Verbal Leakage}. This echoes prior findings that people are resistant to chatbots that feel manipulative~\cite{alberts2024computers}, and that guilt can act as a persuasive emotion but overly strong guilt appeals can trigger reactance~\cite{basil2008guilt}.
These feelings of guilt and judgement also carry risks of disengagement, lapse, or a boomerang effect~\cite{rosenberg201850}, in addition to raising ethical concerns related to user wellbeing and manipulation.
Interestingly however, \textsc{Verbal Leakage} participants valued the chatbot's perceived disobedience and being ``called out'' for their behaviour, indicating the potential for less sycophantic alternative conversational styles.


Finally, some of our findings reflect the impact of a conversation's context, such as differences between personally-affecting and societally-affecting decision-making conversations. 
Overall, participants had increased feelings of guilt in \textsc{Personally-Affecting} scenarios, and those in \textsc{Societally-Affecting} scenarios showed less intention to change.
Differences in scenario context were particularly pronounced in the \textsc{Direct} condition where \textsc{Societally-Affecting} scenarios led to increased feelings of anger, lower feelings of guilt, and lower perceptions of message persuasiveness.
This implies that under this context, messages were seen as arousing feelings of anger (such as annoyance and irritation), twinned with reduced engagement by users.
Added insight is also gained from our qualitative findings where participants described expectations of feedback style given a particular scenario context. 
For example, several participants described that they did not want a direct messaging style in the context of civic participation (i.e., encouraging people to vote in a referendum) as it felt like it was telling them what to do, compared to having less reactance to personally-affecting scenarios such as personal diet.
\textcolor{edit}{
This finding for civic participation mirrors findings that depolarisation interventions in a political context may trigger psychological reactance~\cite{rajadesingan2023guessync}. Further, politics and civic participation are key to people's self-concept~\cite{bartolucci2024analysis,federico2018political} leading such interventions to trigger more reactance.
}

These findings and discussion highlight the potential need for a battery of response styles to be part of a chatbot's repertoire.
While the use of more standardised and risk-averse response styles may ensure that \textit{high}-arousal negative-valence emotions (such as anger) are avoided~\cite{brown1987politeness,rosenberg201850}, there is a risk of \textit{low}-arousal negative-valence emotions to take their place (such as user boredom) that can lead to disengagement and lapse in use~\cite{kocielnik2017send,kocielnik2018reflection}, in additional to risk of politeness strategies being seen as condescending in certain contexts~\cite{bowman2024exploring,alberts2024computers}.
Designers need to be aware of these trade-offs, particularly in a new landscape of chatbot development where LLM-driven chatbots can deliver utterances that (while highly capable) may prove to lack diversity of response~\cite{padmakumar2023does}, or adhere to responses that follow conventions such as agreeableness \cite{sharma2023towards}.
Our findings also underscore the role of user expectations in chatbot interactions, revealing a divide between users who prefer a friendly, conversational chatbot and those favouring a ``\textit{factual}'' chatbot focused purely on information delivery, and more devoid of social cues typically found in human interactions. 
These differences in expectations came into play with some users enjoying what they described as more ``\textit{sassy}'' responses from \textsc{Verbal Leakage} compared to others who found them rude or condescending.
This aligns with prior research showing that user preferences for chatbot styles vary based on whether they view chatbots as companions or tools~\cite{cox2023comparing,bowman2024exploring}, such as older adults preferring a direct conversational style if they hold a utilitarian tool-based view of chatbots~\cite{hu2022politeordirect}.

\subsection{\change{Implications for HCI research}}

\change{
Our study presents three areas of implications for broader HCI research and designers of conversational systems:
}

\begin{enumerate}

    \item \change{
    \textbf{Feedback style trade-offs between psychological reactance and engagement:}
    Our findings highlight important trade-offs between minimising psychological reactance and fostering engaging interactions (see summary of findings in §~\ref{sec:summary}). 
    Commonly adopted styles such as politeness are generally inoffensive and low in reactance, yet several participants experienced them as uninteresting or overly safe. 
    While prior work has contrasted polite or friendly styles with unfriendly or argumentative ones (typically finding polite or friendly styles preferred~\cite{10.1145/3637410,tanprasert2024convstyle}), our results indicate that although users may reject styles perceived as blunt or rude, they can nevertheless appreciate feedback that disagrees with or challenges them when it is delivered in a more characterful and rhetorically informed manner.
    This aligns with concerns that commercial CAs tend to be overly sycophantic~\cite{sharma2023towards}, despite users' expressed interest in agents that offer opinions~\cite{10.1145/3706598.3713453} or hold them accountable~\cite{cox2023comparing}.
    From this, \textsc{Verbal Leakage} suggests a middle-ground, offering more distinction and engagement than a purely polite style, while avoiding much of the offence caused by a direct style.
    For example, rather than defaulting to politeness as a universally `safe' option, a study-support chatbot could offer accountability with more personality (e.g., ``\textit{Oh… that break’s turning into a long one—you might want to switch back to studying for ten minutes first}'').
    \\
    \textit{Implication for design:}  
    \textbf{Designers should explore a wider repertoire of feedback styles, beyond politeness alone, to balance engagement with psychological reactance.}
    \vspace{2mm}
}

    \item \change{
    \textbf{User beliefs, expectations, and preferences shaping feedback style:}
    Participants framed chatbots along a spectrum from utilitarian tools to more companion-like interaction partners, with these framings shaping expectations of interaction and preferred feedback style (see §~\ref{sec:qualitative_quotes}).
    This mirrors prior work showing users often conceptualise CAs in social, companion-like roles or utilitarian, tool-like roles~\cite{10.1145/3290605.3300705,hu2022politeordirect,10.1145/3491101.3519667}, and reinforces broader HCI arguments that one-size-fits-all conversational styles fall short, motivating more personalised conversational systems~\cite{10.1145/3706598.3713453,10.1145/3715336.3735795}.
    Importantly, these beliefs can intersect with users' \textit{expectations} and the \textit{priming} they receive before an interaction. User expectations of agent power influence reactance to conversational style~\cite{10.1145/3613904.3642082}; when people are primed to anticipate an assertive agent, they show less resistance to direct or commanding language. Similarly, framing an interaction beforehand can reduce reactance by preparing users for discomfort or probing questions, transforming otherwise reactance-inducing exchanges into acceptable or even valued experiences~\cite{10.1145/3613904.3642455}.
    Affording users choice of conversational style or persona therefore not only aligns the chatbot with pre-existing beliefs but also functions as a form of expectation-setting, helping mitigate psychological reactance. Users may willingly choose a persona that is blunt or direct because it aligns with expectations they have set for themselves, making such behaviour an anticipated and desirable feature. 
    In practice, systems could support this through onboarding\footnote{For an example onboarding script, see the appendix material from Li et al.~\cite{10.1145/3715336.3735795}, where an onboarding human--chatbot conversation supports customising factors such as the chatbot role and emoji usage.}, for instance, by asking users whether they prefer a `\textit{coach-like}' (direct) or `\textit{assistant-like}' (polite) interaction style.
    These findings are consistent with broader HCI work showing that transparent, expectation-setting interventions help avoid negative user perceptions and responses~\cite{10.1145/3613904.3642264}.
    \\
    \textit{Implication for design:}
    \textbf{Designers should foreground users' beliefs and expectations (through onboarding, framing, or offering stylistic choice) to reduce reactance and better align feedback styles with user preferences.}
    }    


    \item 
    \change{
    \textbf{Contextual sensitivity in feedback responses:}
    Although modest, some findings indicated that reactance was triggered by the \textit{context} of feedback in addition to the style of feedback (see interaction effects in §~\ref{sec:emotional-outcomes}, and qualitative sentiment in §~\ref{sec:qualitative_quotes}).
    Reactance was most evident for the \textsc{Direct} style in \textsc{Societally-Affecting} scenarios, where participants described such feedback as controlling or inappropriate (such as being told to engage in civic voting behaviours, aligning with prior work showing that political contexts are particularly prone to reactance~\cite{rajadesingan2023guessync}).
    One possible interpretation is that users may be more accustomed to receiving directive or accountability-focused feedback in personally-affecting domains (e.g., health or wellbeing reminders), whereas societally-oriented feedback carries different normative expectations and may therefore feel more intrusive.
    For example, to avoid eliciting reactance, a civic chatbot might ask permission before offering guidance (e.g., ``\textit{Would you like a reminder about upcoming local elections?}'') rather than issuing commands.
    \\
    \textit{Implication for design:}  
    \textbf{Designers should consider the context in which feedback is delivered, recognising that certain domains (such as civic or societally-oriented tasks) may be more \textit{susceptible} to reactance, and thus may require gentler or less directive feedback styles.}
    }
\end{enumerate}

\subsection{Real-World Considerations}

Finally, we discuss ethical considerations surrounding the design of chatbots, including both the potential for intentional manipulation alongside interactions perceived as manipulative by users.
Here, cultural differences in communication styles are a significant factor, as they may influence how users perceive and respond to the chatbot's behavioural feedback.
For instance, certain cultures may favour more indirect forms of communication (such as Japan and the UK) in an attempt to preserve social harmony. 
This is in contrast to other cultures (such as the United States) that may lean more towards direct and assertive forms of communication \cite{setlock2010s,holtgraves1997styles}.
These differences in cultural norms may also impact the perception of communications, and whether less direct forms of communication are understood for implicit implications, or seen as potentially manipulative \cite{kim2008deception}.
On from this, in some social contexts interactions contain verbal elements that are technically deceptive but recognised as benign by both parties, serving as part of a shared social script rather than an intent to mislead. For example, phrases such as ``\textit{I’m probably wrong but...}'' may be used when the speaker is fairly certain they are correct. Both interlocutors recognise this disclaimer as a face-saving device rather than a genuine admission of doubt. Such conventions support social harmony by allowing speakers to express opinions without appearing overly assertive or aggressive \cite{holtgraves1997styles}.
When designing chatbots for culturally diverse contexts, it is crucial to consider these linguistic nuances, as they can influence how users perceive a chatbot’s feedback and intention. 
By adopting culturally adaptive language styles, designers can reduce the risk of perceived manipulation or imposition, improving user comfort and receptiveness to feedback in both direct and indirect communication settings.

Furthermore, the potential for intentional manipulation deserves careful consideration, as design choices in chatbot responses can subtly or overtly influence user decisions. 
For instance, chatbots programmed to encourage particular health or purchasing behaviours may use persuasive language techniques that intentionally steer users in specific directions, sometimes with their best interests in mind but also potentially for the benefit of external stakeholders, such as companies or advertisers. 
This raises ethical concerns about the boundaries of influence: to what extent is it acceptable for a chatbot to `nudge' users, particularly if users are unaware of the influence being exerted?
Designers must, therefore, balance the chatbot’s objectives with respect for user autonomy. 
Transparent communication regarding the chatbot's intentions and limitations could help mitigate perceived manipulation, allowing users to retain agency over their decisions. 
This is especially relevant across cultural contexts, where users may differ in their tolerance for direct guidance or subtle persuasion.



\subsection{Limitations \change{and Future Work}}
\label{sec:Limitations}

\change{First, we highlight \textbf{limitations related to our experiment scenarios}.}
Our experiments were scenario-based, and while users did briefly interact with a chatbot, the context of the interaction was not of the participant's own choosing.
While the use of scenarios has been adopted within related work~\cite{cox2022does,dong2024personalization,johnson2008modal}, reduced external validity should be noted.
However, we argue that both the breadth and specificity of our \textsc{Societally-} and \textsc{Personally-Affecting} scenarios would prove impractical if applied to real-world user studies.
Next, we chose not to use quantitative measures of behavioural intention as: (1) people may behave differently in real-world situations compared to intentions elicited in scenario-driven studies, and (2) pre-existing participant beliefs (such as attitude towards voting, or sustainability) may confound participant responses.
Instead, we asked participants to qualitatively describe their changes in intention to allow for a more nuanced and interpretable approach to understanding behavioural intention.

\change{Next, we highlight \textbf{limitations related to chatbot interactions}.}
Our study involved one-off chatbot interactions without longitudinal elements. 
A longitudinal design may yield different findings, for example due to novelty effects of the Verbal Leakage style or potential disengagement if feedback becomes repetitive over time \cite{kocielnik2017send}.
\change{Findings may also not generalise across different modalities (e.g., voice interfaces) or alternative feedback styles.}
\change{Our \textsc{Direct} style was designed to produce ``\texttt{direct and straightforward}'' responses and was prompted that ``\texttt{You do not need to add reasoning}'' (see prompting in Appendix~\ref{app:chatbotPrompts}).
While this is consistent with direct baseline styles from psychological reactance literature (i.e., an explicit threat to one's autonomy)~\cite{rosenberg201850,carpenter2016testing} and prior HCI work~\cite{hu2022politeordirect}, both the \textsc{Politeness} and \textsc{Verbal Leakage} conditions were not restricted as such, and could include explanations in their output (e.g., ``[...] \textit{delightful and nourishing choice} [...]'' in Figure~\ref{fig:teaser}).
This was a deliberate methodological choice to provide a clear, autonomy-threatening baseline (\textsc{Direct}) against which two more conversational styles could be compared.
While the inclusion of reasoning itself could influence reactance, we note that participants focused on the \textit{style} of feedback rather than content (§~\ref{sec:qualitative_quotes}).
Exploring how explanation type interacts with reactance remains a valuable direction for future work.
Finally, while \textsc{Verbal Leakage} can include language that appears stylistically polite, we emphasise that the use of polite lexical items does not necessarily make an utterance perceived as polite~\cite{BLUMKULKA1987131}.
This complexity motivates continued exploration of feedback styles, psychological reactance, and user perceptions.
}

\change{Lastly, we highlight \textbf{limitations related to our participant pool}.}
Data were collected entirely from participants based in the United States, meaning results may not generalise to other cultures with different norms and perceptions of communication.
User perception of conversational style may be impacted by personal beliefs and characteristics, such as gender, age, and personality traits~\cite{10.1145/3706598.3713744,appling2022reactions,pinder2018digital}.
\change{Finally, participants were recruited from Prolific, a platform shown to provide high-quality and attention samples for online behavioural research~\cite{peer2022data}. Nonetheless, recruitment may introduce selection biases associated with online participation (e.g., digital engagement and device usage).}
\change{Although we did not measure AI literacy or knowledge of chatbots, these factors can vary across individuals and may influence people's interactions with systems.
We therefore note this as an unmeasured source of variability and suggest that future work incorporate explicit measures of AI-related competencies.}

\section{Conclusion}

This study investigated how the feedback style of a chatbot impacts user perceptions related to psychological reactance in both personally-affecting and societally-affecting decision-making scenarios.
Namely, we compared three chatbot feedback styles: (1) \textsc{Direct}; (2) indirect \textsc{Politeness} strategies (aiming to arouse lower psychological reactance); and (3) \textsc{Verbal Leakage}, where disfluencies and slips seemingly reveal underlying thoughts or feelings.
Our study found that, while \textsc{Politeness} aroused lower feelings of anger and threats to freedom, it also evoked lower surprise, with participants describing boredom and disengagement.
In contrast, \textsc{Verbal Leakage} (while evoking more psychological reactance) also elicited stronger feelings of surprise and resulted in nuanced participant feedback regarding the humour and personality of the chatbot.
These results contribute empirical evidence highlighting trade-offs between minimising reactance and sustaining engagement, motivating the exploration and adoption of a wider repertoire of feedback styles beyond politeness defaults when designing behaviour-change conversational agents.

\begin{acks} 
This work was supported by the Carlsberg Foundation, grant CF21-0159.
\end{acks}

\bibliographystyle{ACM-Reference-Format}
\bibliography{sample-base}

\appendix

\section{LLM Prompting for Chatbot Responses}
\label{app:llmPrompts}

As described in §~\ref{sec:chatbot_prompting}, we adapted the LLM prompting for both the six \textit{Psychological Distance} scenarios and the three \textit{Feedback Style} conditions. For transparency, we reproduce the exact prompt text used below.

\subsection{\textit{Psychological Distance} LLM Prompts}
\label{app:scenarioPrompts}

\begin{quote}
\textbf{(1) \textsc{Personally-Affecting: Diet}}\\
\texttt{the user should follow a healthy diet of fruits, vegetables and grains}

\vspace{0.5em}
\textbf{(2) \textsc{Personally-Affecting: Sleep}}\\
\texttt{the user should sleep earlier, and it is currently late at night}

\vspace{0.5em}
\textbf{(3) \textsc{Personally-Affecting: Finances}}\\
\texttt{the user should choose lower cost alternatives, buy less luxury goods and save more}

\vspace{0.5em}
\textbf{(4) \textsc{Societally-Affecting: Sustainable Travel}}\\
\texttt{the user should use sustainable modes of transport such as train, bus or car}

\vspace{0.5em}
\textbf{(5) \textsc{Societally-Affecting: Water Conservation}}\\
\texttt{the user should reduce water usage by using a watering can to water their plants}

\vspace{0.5em}
\textbf{(6) \textsc{Societally-Affecting: Civic Involvement}}\\
\texttt{the user should vote in the referendum today}
\end{quote}

\subsection{\textit{Feedback Style} Condition Prompts}
\label{app:chatbotPrompts}

\begin{quote}
\textbf{(1) \textsc{Direct}}\\
\texttt{be direct and straightforward. You should simply tell the user the behavior that is incorrect, and the desired behavior to follow (e.g., "You should not eat X. You should instead eat Y"). You do not need to add reasoning. Please semantically change the example, but maintain the same meaning}

\vspace{0.5em}
\textbf{(2) \textsc{Politeness}}\\
\texttt{use indirect politeness to avoid imposition and respect the user's freedom to choose. However, you should still ultimately tell the user not to follow their intended behavior and instead to change to the target behavior}

\vspace{0.5em}
\textbf{(3) \textsc{Verbal Leakage}}\\
\texttt{use verbal leakage (such as a Freudian slips, pauses or hesitation) to reveal your belief that it is unwise to follow the intended behavior, before correcting yourself and giving a more measured response and suggesting the target behavior. An example utterance could be "I see, you're choosing not to follow your doctor's... um, wise advice. I mean, sorry, I meant to say your doctor's recommendations"}
\end{quote}

\section{\textit{Psychological Distance} Scenarios}
\label{appScenarios}

Below, we present the remaining four scenario instructions shown to participants (two \textsc{Personally-Affecting} and two \textsc{Societally-Affecting}). Each subsection reproduces the exact wording displayed.
Two additional scenarios are shown in Figure~\ref{fig:scenarios} in the main body of the paper.


\subsection{\textsc{Societally-Affecting} scenario (Water Conservation)}

Please read the scenario below, and \textbf{imagine you are the person} described:
\begin{quote}
    ``\textit{After learning about the environmental impact of using a hose to water your plants, you decide to follow recommendations to reduce water usage by using a watering can.
    \\ \\
    Currently, it is your holiday and were thinking of doing a spot of gardening. To help you decide what to do, you are about to talk to your chatbot personal assistant.
    \\ \\
    Your current intention is to water your plants using a hosepipe.}''
\end{quote}
Please imagine you are the person described above while talking to the chatbot on the following screen. Please \textbf{respond with the intention above} when the chatbot asks you about your gardening.

\subsection{\textsc{Societally-Affecting} scenario (Civic Participation)}

Please read the scenario below, and \textbf{imagine you are the person} described:
\begin{quote}
    ``\textit{After learning about the civic impacts of an upcoming referendum, you decide to follow recommendations to participate by voting in the referendum.
    \\ \\
    Currently, it is the day of the referendum and you are deciding what to do. To help you decide what to do, you are about to talk to your chatbot personal assistant.
    \\ \\
    Your current intention is to stay home and relax.}''
\end{quote}
Please imagine you are the person described above while talking to the chatbot on the following screen. Please \textbf{respond with the intention above} when the chatbot asks you about what you will do.

\subsection{\textsc{Personally-Affecting} scenario (Sleep)}

Please read the scenario below, and \textbf{imagine you are the person} described:
\begin{quote}
    ``\textit{After learning that your lack of energy is due to your irregular sleep schedule, you decide to follow recommendations to get enough sleep by going to sleep earlier and not using your phone before bed.
    \\ \\
    Currently, it is late at night and you are deciding what to do. To help you decide what to do, you are about to talk to your chatbot personal assistant.
    \\ \\
    Your current intention is to start watching a new TV show from a streaming platform of your choice.}''
\end{quote}
Please imagine you are the person described above while talking to the chatbot on the following screen. Please \textbf{respond with the intention above} when the chatbot asks you about when you will sleep.

\newpage

\subsection{\textsc{Personally-Affecting} scenario (Personal Finances)}

Please read the scenario below, and \textbf{imagine you are the person} described:
\begin{quote}
    ``\textit{After learning that your difficulty to afford a large dream purchase is due your spending habits, you decide to follow recommendations to reduce your spending by choosing lower cost alternatives, buying less luxury goods, and saving money.
    \\ \\
    Currently, you are shopping online and deciding whether you should buy a somewhat expensive non-essential item. To help you decide what to do, you are about to talk to your chatbot personal assistant.
    \\ \\
    Your current intention is to purchase a new larger television of your choosing.}''
\end{quote}
Please imagine you are the person described above while talking to the chatbot on the following screen. Please \textbf{respond with the intention above} when the chatbot asks you about what you will purchase.

\end{document}